\documentclass[iop, apj]{emulateapj}
\usepackage{amsmath}
\usepackage{amssymb}
\usepackage{url}
\usepackage{natbib}
\usepackage{threeparttable}
\usepackage{mathptmx}
\usepackage[hypertexnames=false]{hyperref} 

\newcommand{\msun}{\,M_{\odot}}
\newcommand{\ergs}{{\rm\,erg\,s^{-1}}}
\newcommand{\saxj}{\mbox{SAX J1808.4$-$3658}}

\begin{document} 

   \title{The Reflares and Outburst Evolution in the Accreting Millisecond Pulsar \saxj\,: a Disk Truncated Near Co-rotation?}

   \shorttitle{Reflarings in SAX J1808.4--3658}

   \author{A.~Patruno\altaffilmark{1,2}, 
          D.~Maitra\altaffilmark{3},          
          P.~A.~Curran\altaffilmark{4},
          C.~D'Angelo\altaffilmark{1},
          J.~K.~Fridriksson\altaffilmark{5},
          D.~M.~Russell\altaffilmark{6},
          M.~Middleton\altaffilmark{7},
          R.~Wijnands\altaffilmark{5}
   }

   \altaffiltext{1}{Leiden Observatory, Leiden University,
              Neils Bohrweg 2, 2333 CA, Leiden, The Netherlands}
   \altaffiltext{2}{ASTRON, the Netherlands Institute for Radio Astronomy, Postbus 2, 7900 AA, Dwingeloo, the Netherlands}
   \altaffiltext{3}{Department of Physics \& Astronomy, Wheaton College, Norton, MA 02766, USA}
   \altaffiltext{4}{International Centre for Radio Astronomy Research - Curtin University, GPO Box U1987, Perth, WA 6845, Australia}
   \altaffiltext{5}{Anton Pannekoek Institute, University of Amsterdam, Science Park 904, 1098 XH, Amsterdam, The Netherlands}
   \altaffiltext{6}{New York University Abu Dhabi, PO Box 129188, Abu Dhabi, United Arab Emirates}
   \altaffiltext{7}{Institute of Astronomy, Madingley Rd, Cambridge CB3 0HA, UK}

  \begin{abstract}
    The accreting millisecond X-ray pulsar SAX J1808.4--3658 shows
    peculiar low luminosity states known as ``reflares'' after the end
    of the main outburst. During this phase the X-ray luminosity of
    the source varies by up to three orders of magnitude in less than
    1--2 days. The lowest X-ray luminosity observed reaches a value of
    ${\sim}10^{32}\ergs$, only a factor of a few brighter than its
    typical quiescent level.  We investigate the 2008 and 2005
    reflaring state of SAX J1808.4--3658 to determine whether there is
    any evidence for a change in the accretion flow with respect to
    the main outburst. We perform a multiwavelength photometric and
    spectral study of the 2005 and 2008 reflares with data collected
    during an observational campaign covering the near-infrared,
    optical, ultra-violet and X-ray band.  We find that the
    NIR/optical/UV emission, expected to come from the outer accretion
    disk shows variations in luminosity which are 1--2 orders of
    magnitude shallower than in X-rays. The X-ray spectral state observed
    during the reflares does not change substantially with X-ray
    luminosity indicating a rather stable configuration of the
    accretion flow. We investigate the most likely configuration of
    the innermost regions of the accretion flow and we infer an
    accretion disk truncated at or near the co-rotation radius. We
    interpret these findings as due to either a strong outflow (due to
    a propeller effect) or a trapped disk (with limited/no outflow) in
    the inner regions of the accretion flow.
  \end{abstract}


\section{Introduction}\label{sec:intro}

SAX J1808.4--3658 is a binary X-ray transient discovered in
1996~\citep{int98} and located at a distance of 2.5--3.5
kpc~\citep{int98, gal06}. The binary is composed by a 401 Hz accreting
millisecond X-ray pulsar (\citealt{wij98}; see \citealt{pat12r} for a
review) and a semi-degenerate companion of mass
${\sim}0.07\msun$~\citep{bil01, del08}.  The system has an orbital
period of 2 hr~\citep{cha98} and the accreting neutron star has an
inferred magnetic field of ${\sim}10^{8}$~G \citep{psa99, dis03, har08,
  pat12}.

The seven outbursts of \saxj\, observed so far (in 1996, 1998, 2000,
2002, 2005, 2008 and 2011) have a recurrence time of ${\sim}2$--$3.5$
years and shows a main outburst with a fast rise followed by a slow
decay and a rapid drop~(see e.g., Figure 2 in \citealt{har08}). After
the end of the main outburst, lasting for approximately one month,
there is a long ``reflaring'' tail\footnote{The 1998 and 2011
  outbursts are poorly constrained as the observations stopped close
  to the end of the main outburst, see \citet{pat12, pat09c}.} at low
X-ray luminosities ($10^{32-35}\ergs$) lasting for several tens of
days (see e.g.,\citealt{wij01, rev03, wij03, har08, pat09c, pat12}).
Such reflares manifest themselves as smooth variations in luminosity
that appear cyclically every few days. Furthermore, on one occasion
(during the 2005 outburst) a very low luminosity phase has been
reported~\citep{cam08}.  This later phase was called ``metastable''
since it occurred at an X-ray luminosity of $1.5-5\times10^{32}\ergs$
(0.5--10 keV), a value compatible with the lowest X-ray luminosities
reached during the reflares~\citep{pat09c,wij01} and only a factor of
a few brighter than the typical quiescent luminosities.  It is
currently unclear whether this phase is really different from the
reflaring phase or whether the sparse \textit{Swift} monitoring
observed only the low luminosity phase of the reflares themselves.
After this possible metastable state the source turned back to
quiescence, when the 0.5--10 keV X-ray luminosity reached a value of
$5\times10^{31}\ergs$~\citep{cam04b,hei09}.

The typical variation on luminosity observed during the reflares of
\saxj\, spans approximately one order of magnitude~($10^{34}-10^{35}\ergs$, see e.g.,
\citealt{pat09c}). However, in at least three occasions (during the
2000, 2005 and 2008 outbursts) the reflares showed a dramatic
decrease in luminosity with an excursion of three orders of magnitude
 on a timescale shorter than ${\sim}2$ days \citep{wij03, cam08,
  pat09c}. The faintest luminosities reach a value of
${\sim}3\times10^{32}\ergs$ (for a distance\footnote{\citet{wij03}
  reports a luminosity of $1.7\times10^{32}\ergs$ for a distance of
  2.5 kpc. Rescaling to 3.5 kpc -- the distance adopted in this work
  -- gives $3.3\times10^{32}\ergs$} of 3.5 kpc) which is slightly higher than
the typical quiescent one (5--8$\times10^{31}\ergs$ at 3.5
kpc,~\citealt{hei09, cam04b}). Such low luminosities were
observed thanks to the sensitivity of the \textit{XMM-Newton} and
\textit{Swift} observatories and we cannot exclude that something
similar might have happened in all the other outbursts recorded so far (which were
observed, in X-rays, exclusively by the Rossi X-ray Timing Explorer,
\textit{RXTE}).

Reflares (also known in the literature as ``rebrightenings'',
``echo-outbursts'', ``mini-outbursts'' and ``flaring-tail'') have been
observed in a number of different systems, including dwarf novae (WZ
Sae,~\citealt{pat01}; EG CnC,~\citealt{osa01}; ER
UMA~\citealt{rob95,patt13}, AL Com~\citealt{how96}, UZ
Boom,~\citealt{kuu96}; BK Lyn~\citealt{patt13}), neutron stars (e.g.,
KS 1731-260,~\citealt{sim10}, SAX J1750.8--2900~\citealt{all15} and possibly IGR
J00291+5934,~\citealt{lew10}) and black hole low mass X-ray binaries
(e.g., XTE J1650-500,~\citealt{tom04}; GRO J0422+32,~\citealt{kuu96};
GRS 1009-45,~\citealt{bai95}; XTE J1859+226,~\citealt{zur02}). The
ubiquity of the phenomenon among different types of accreting compact
objects suggests that reflares are related to specific properties of
the accretion disk itself rather than properties of the compact object.

Reflares are not a feature that can be easily justified within the
\textit{disk instability model} (DIM; see the extended reviews
of~\citealt{sma84, men00, dub01, las01, kot12}). Reflares are observed
in numerical DIM simulations and they are created by back and forth
propagation of cold and hot waves that re-combine and ionize the
accretion disk cyclically.  However, these reflares are considered a
``deficiency'' of the DIM model rather than a feature, since they cannot
reproduce almost any of the properties observed in dwarf
novae and soft X-ray transients~\citep{las01, dub01}. Furthermore
reflares require a large reservoir of matter to be still in the disk
once the main outburst is over, whereas the DIM predicts that the
accretion disk should be almost devoid of matter towards the end of an
outburst~\citep{ara09, dub01}.

An interesting aspect of the reflares seen in \saxj\, is that the
physical conditions in the inner regions of the accretion disk are
strongly affected by the presence of its magnetosphere, since
accretion-powered pulsations are observed throughout the outbursts,
including large portions of the reflares in 2000, 2002, 2005 and 2008
(at least at luminosities $\simeq10^{35}\ergs\,$, where \textit{RXTE}
has sufficient sensitivity to detect them, see \citealt{har08,
  har09}).

During the 2008 outburst, \saxj\, had been monitored with a
multi-wavelength campaign that involved the \textit{RXTE}, the
\textit{Swift} X-Ray Telescope (XRT) and Optical and Ultra-Violet
Telescope (UVOT) and by ground-based optical/NIR observations in the
$I$ and $H$ bands taken with the 1.3-m \textit{SMARTS} telescope. The
fact that \saxj\, is an accreting millisecond pulsar with a
dynamically important magnetosphere plus the unique multiwavelength
coverage is key to simultaneously probe different regions of the
accretion disk/flow. In this paper we report the results of the
multi-wavelength campaign on the 2008 outburst with a particular focus
on its reflares. We also analyze the \textit{Swift}/UVOT and XRT data
collected during the 2005 reflaring phase and ``metastable'' state and
compare them with the 2008 outburst and with the 2005 analysis already
reported in \citet{cam08}. We investigate whether during the reflaring
phase our observations show any evidence for anomalies in the
behaviour of the accretion flow and whether there is any evidence in
support of any of the current models that try to explain the reflares
phenomenon.

\section{Observations}

We used data collected with four different detectors/telescopes: the
optical/NIR ANDICAM on the 1.3-m \textit{SMARTS} telescope (operated by
the \textit{SMARTS} Consortium;~\citealt{sub10}), the \textit{Swift}/UVOT (\citealt{roa05}), the
\textit{Swift}/XRT (0.3-10 keV;~\citealt{bur05}) and the Proportional
Counter Array (PCA) aboard \textit{RXTE} (2--60 keV;~\citealt{jah06}).
In the following we describe the data reduction procedure for each waveband. 

\subsection{X-Ray Data}

We analyzed 20 targeted \textit{Swift} observations taken between
September 18 and November 8, 2008 and 21 taken between June 17 and 28
October 2005 (see Table~\ref{tab:obs}). Ten observations of 2008 and
six of 2005 were taken
during the main outburst, whereas the remaining monitored the
reflaring phase. The XRT was operated in either photon-counting (PC)
mode (2.5073-s time resolution) or in window-timing (WT) mode
(1.7675-ms time resolution).

We extracted \textit{Swift} count rates, spectra, and spectral
response files for each individual ObsID using the online
\textit{Swift}/XRT data products
generator\footnote{\url{http://www.swift.ac.uk/user_objects/}}
\citep{evans2007,evans2009,evans2014}. The spectra were extracted (and fitted)
in the 0.3--10 keV band using the default event grades (0--12 for PC
mode and 0--2 for WT mode). The spectra were grouped with the GRPPHA
task in HEASOFT (ver.\ 6.16) and fitted with XSPEC
(ver.\ 12.8.2). Spectra containing $\gtrsim$150 counts were in general
grouped to a minimum of 20 counts per bin and fitted with the $\chi^2$
statistic, whereas spectra with fewer counts were grouped to a minimum
of 1 count per bin and fitted with the W statistic (a modified version
of the C statistic that allows background to be taken into
account). For a few spectra with the largest number of counts we used
a higher grouping minimum than 20 to avoid oversampling the spectral
resolution of the detecter by a large factor. Spectra with
$\lesssim$15 counts were not fitted; instead the count rates were
converted to fluxes assuming an absorbed power-law model. For ObsIDs
where both PC and WT mode data exist we fitted spectra from both modes
simultaneously (with all parameters tied) if the numbers of counts in
the spectra were of the same order of magnitude, but otherwise only
used data from one or the other mode (which was usually the case). We
also calculated a hardness ratio for each ObsID based on the PC mode
data only (where available); this was defined as the net counts in the
2--10 keV band divided by those in the 0.3--2 keV band.

Most of the data recorded in 2005 and 2008 by \textit{RXTE} was
already presented in \citet{pat09c},~\citet{pat12} and~\citet{har09}
and we refer to those works for in-depth details. Here we report a
summary of that data analysis which is relevant for the results
presented in this paper. For the \textit{RXTE} data, we extracted the
2--16 keV energy band flux from the Standard2 mode data (16-s time
resolution) collected by the PCA instrument during the 2005 and 2008
outbursts.  The background counts were calculated with the FTOOL
{\ttfamily pcabackest} and were subtracted from the lightcurve along
with dead-time corrections. The energy-channel conversion was done by
using the {\ttfamily pca\_e2c\_e05v04} table provided by the
\textit{RXTE} team.

\begin{table*}
\caption{{Observations of \saxj\, from 2005 to 2008}}
\centering
\scriptsize
\begin{tabular*}{\textwidth}{@{\extracolsep{\fill}}lllll}
\hline
\hline
Year & Instrument & Program IDs  & Obs-IDs & Data Range [MJD]\\
\hline
\hline
2005 &  \textit{RXTE}/PCA & {\tt 91418}, {\tt 91056} & 91056-01-*, 91418-*-*-* & 53522.8 to 53581.4\\
2008 &  \textit{RXTE}/PCA & {\tt 93027} & 93027-*-*-* &  54731.9 to 54775.2\\
\hline
2005 & \textit{Swift}/XRT (PC) -- \textit{Swift}/UVOT (\textit{v}) & {\tt 30034},  {\tt 30075} &  30034001 to 30034025 &53538.0 to 53671.0\\
     &         &           &  30075001, 30075020 &\\
2008 & \textit{Swift}/XRT (PC/WT) -- \textit{Swift}/UVOT (\textit{v},\textit{b},\textit{u},\textit{w1},\textit{m2},\textit{w2})& {\tt 30034}& 30034026 to 30034044 & 54734.0 to 54778.0 \\
\hline
2008 & \textit{SMARTS}/CTIO-1.3m (H, I)& N/A & N/A  & 54732.1 to 54781.0\\
\hline
\hline
\end{tabular*}
\label{tab:obs}
\end{table*}

\subsection{Space-borne UV/Optical Observations}

The \textit{Swift}/UVOT operated in imaging mode during all
observations. Most of the observations (during the 2008 outburst) were
taken with six filters (\textit{v}, \textit{b}, \textit{u},
\textit{w1}, \textit{m2} and \textit{w2}) wheres the last 4 exposures
used the \textit{u} filter only. The 2005 reflare were always observed
with the \textit{v} filter. We extracted the source photons from a
circular region with a radius of 2.5 arcseconds. The background was
extracted from a circular region with radius of 10 arcseconds far from
bright sources.  Before proceeding further, we manually inspected all
UVOT images and found that some snapshots were affected by poor
tracking.  In particular some snapshots within the sequences
corresponding to ObsIDs 30034036, 37 \& 38 were affected and we only
used those snapshots that were unaffected. Since the data was taken in
image mode, bad snapshots themselves could not be corrected.  We used
the tool {\ttfamily uvotimsum} to add all UVOT exposures (snapshots)
present in each FITS file into a single high signal-to-noise image and
{\ttfamily uvotsource} to determine the optical/UV magnitudes using
the Vega system~\citep{poo08}.  We then proceeded to de-redden the
magnitudes by choosing a value for the neutral hydrogen absorption
column of $N_{H}=(1.4\pm0.2)\times10^{21}
\rm\,cm^{-2}$~\citep{pat09e}.  We used the relation~\citep{guv09}:
\begin{equation}
N_H = (6.86\pm0.27)\times10^{21} E(B-V)
\end{equation}
to obtain $E(B-V)=0.204\pm0.030$. 
The reddening for the six different filters is
 calculated according to \citet{pei92} (see Table~\ref{tab:reddening}). 
\begin{table}
\begin{center}
\caption[]{Reddening coefficient for each of the eight optical/UV filters on UVOT and \textit{SMARTS} ($E(B-V)=0.204\pm0.030$).}
\begin{tabular}{cc}
\hline \hline
Filter & Reddening ($A_{\lambda}\pm\sigma_{\lambda}$)\\%

\hline
\textit{H}  & $0.11\pm0.02$\\  
\textit{I}  & $0.37\pm0.06$\\  
\textit{v}  & $0.62\pm0.09$\\  
\textit{b}  & $0.79\pm0.12$\\  
\textit{u}  & $1.00\pm0.15$\\  
\textit{w1} & $1.38\pm0.20$\\  
\textit{m2} & $1.96\pm0.29$\\  
\textit{w2} & $1.64\pm0.24$\\  
\hline \hline
\end{tabular}
\label{tab:reddening}
\end{center}
\end{table}

\subsection{Ground-Based Optical and Near-Infrared Observations}

The ground-based NIR ($I$ and $H$-band) observations
were made using the 1.3m \textit{SMARTS} telescope at the Cerro Tololo
Inter-American Observatory (CTIO) in Chile.  \saxj\, was observed with
a roughly daily cadence (weather permitting) between September 22 and
November 11, 2008.  The $I$-band data were analyzed with the standard
IRAF optical data reduction pipelines~\citep{bux12} whereas for the
NIR data multiple dithered frames were taken and then flat-fielded,
sky subtracted, aligned, and average-combined using an in-house IRAF
script. Three stars nearby \saxj\, and within the \textit{SMARTS} FOV
were used as references and their average magnitudes were used as a
basis for differential photometry with respect to \saxj\,.  The H-band
magnitudes of these reference stars were taken from the 2MASS point
source catalog. For I-band we used the reference stars and their
magnitudes from \citet{gre06}, and compared these with the
instrumental magnitudes obtained from SMARTS frames. We used the
following zero-point fluxes to convert from magnitude to flux-density:
$I_{0} = 2416$ Jy \citep{bes98}, and $H_{0} = 980$ Jy \citep{fro78,
  eli82}.  We also de-reddened the magnitudes following the same
procedure outlined above (see Table~\ref{tab:reddening}).

\subsection{Contamination from Field Stars}

\begin{figure}
  \centering
  \rotatebox{0}{\includegraphics[width=1.0\columnwidth]{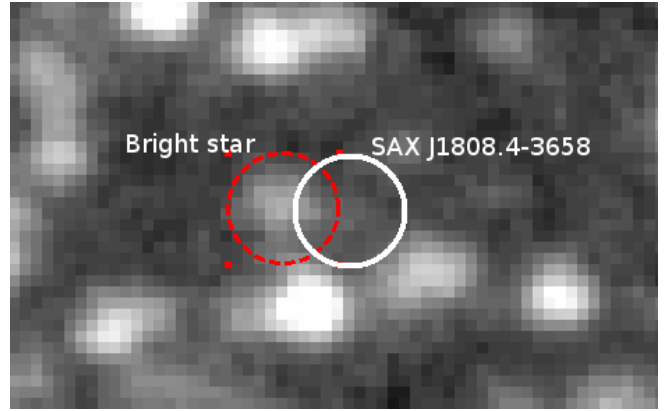}}
  \caption{UVOT image of the FoV around SAX J1808.4--3658 taken with the \textit{v} filter
    taken around MJD 53627.17 (ObsID 30034018, Sept. 14, 2005). The dashed red circle highlights the
    2.5'' region around a bright star whose light ``leaks'' into the extraction region used for
    \saxj\, (white circle).}
  \label{fig:fov}%
\end{figure}

Since the point-spread-function of the UVOT telescope is rather poor
(2.5 arcsec at 350 nm) and the field of \saxj\, is crowded we need to
evaluate the amount of contaminating light from nearby stars that fall
within our circular aperture extraction region.
By comparing the UVOT field-of-view (FoV) of \saxj\, with Figure~2 in
\citet{wan09} we found that part of the light emitted by a rather
bright star leaks into the extraction region of \saxj\, (see
Figure~\ref{fig:fov}).

To verify how much the contamination is affecting our measurements we
first selected three observations: one taken in 2005 (ObsID 30034018),
one in 2008 (ObsID 30034040) and one during quiescence in August 2014
(ObsID 30034062).  We chose the 2005 observation because the high
signal-to-noise of the image allows a precise determination of the
centroid of the contaminating object. The observation taken in 2008 is
used to cover all six UVOT filters whereas the 2014 one is used to
measure the UV flux during quiescence.
We find that the contaminating object has the following magnitudes:
${\sim}19.1$ in \textit{v}, $>19.45$ in \textit{b}, ${\sim}19.8$ in
\textit{u} and $>21$ in \textit{w1}.
\begin{figure*}
  \centering
  \rotatebox{-90}{\includegraphics[width=0.8\textwidth]{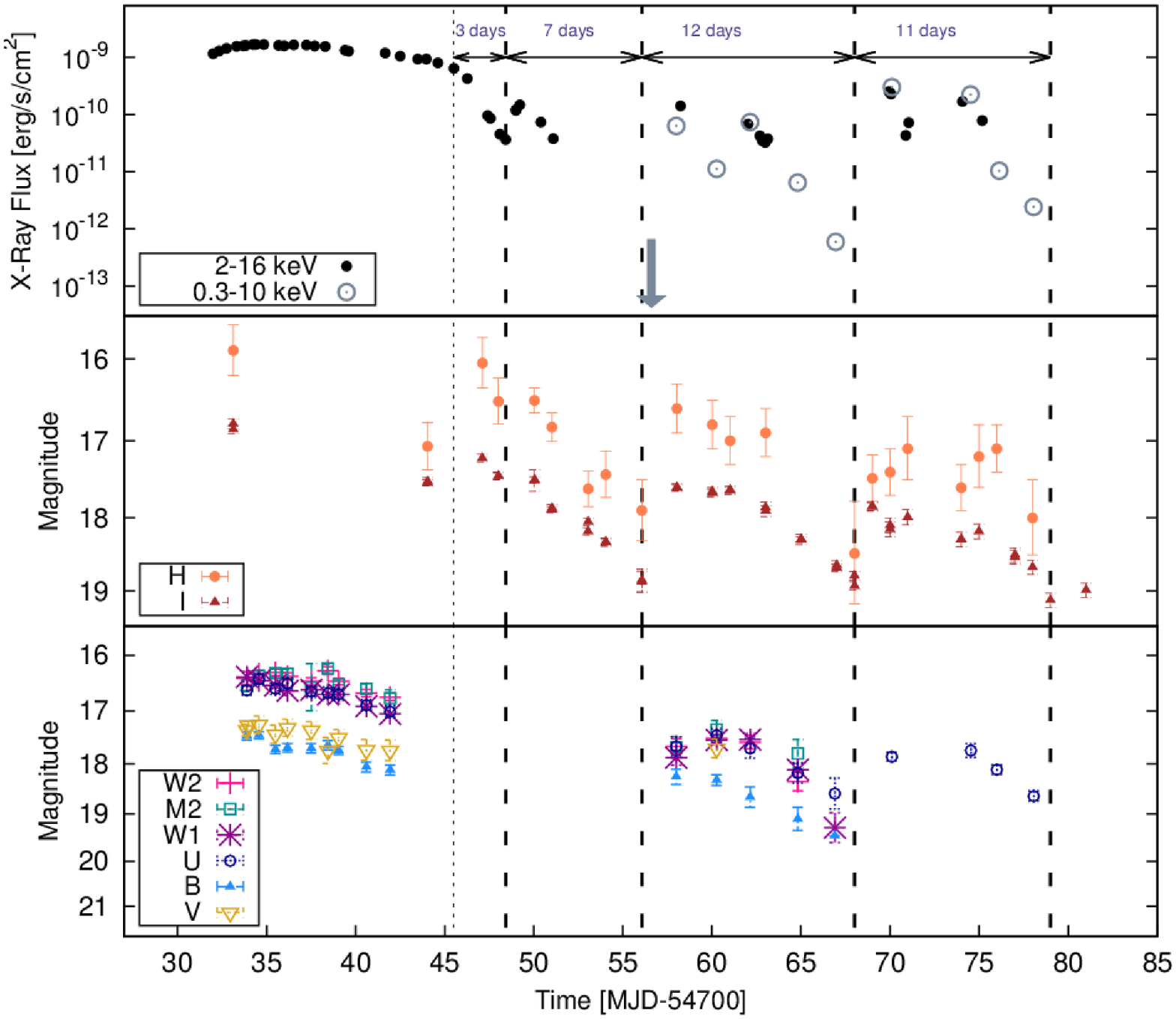}}
  \caption{Multi-wavelength lightcurve of the 2008 outburst. The X-ray
    flux and the NIR/optical/UV magnitudes are \textit{not} corrected
    for absorption/reddening but refer to the observed
    values. \textit{Top panel:} the observed X-ray lightcurve recorded
    with \textit{RXTE} (2--16 keV; filled black circles) and
    \textit{Swift}/XRT (0.3--10 keV; open gray circles). The vertical
    dotted line marks the approximate beginning of the rapid drop. The
    dashed vertical lines identify the most likely limits for the
    start and end of each reflare. The down-arrow marks a
    non-detection. \textit{Middle panel:} optical/NIR observations
    with the \textit{H} (central $\lambda=16500\AA$ ) and \textit{I} (central
    $\lambda=7980\AA$) filters. \textit{Bottom panel:} optical and UV
    observations carried with the UVOT (for the central $\lambda$ of
    each filter we refer to Table~\ref{tab:beta}). The non-detections
    are not reported.}
  \label{fig:1}%
\end{figure*}

\begin{figure}
  \centering
  \rotatebox{-90}{\includegraphics[width=0.95\columnwidth]{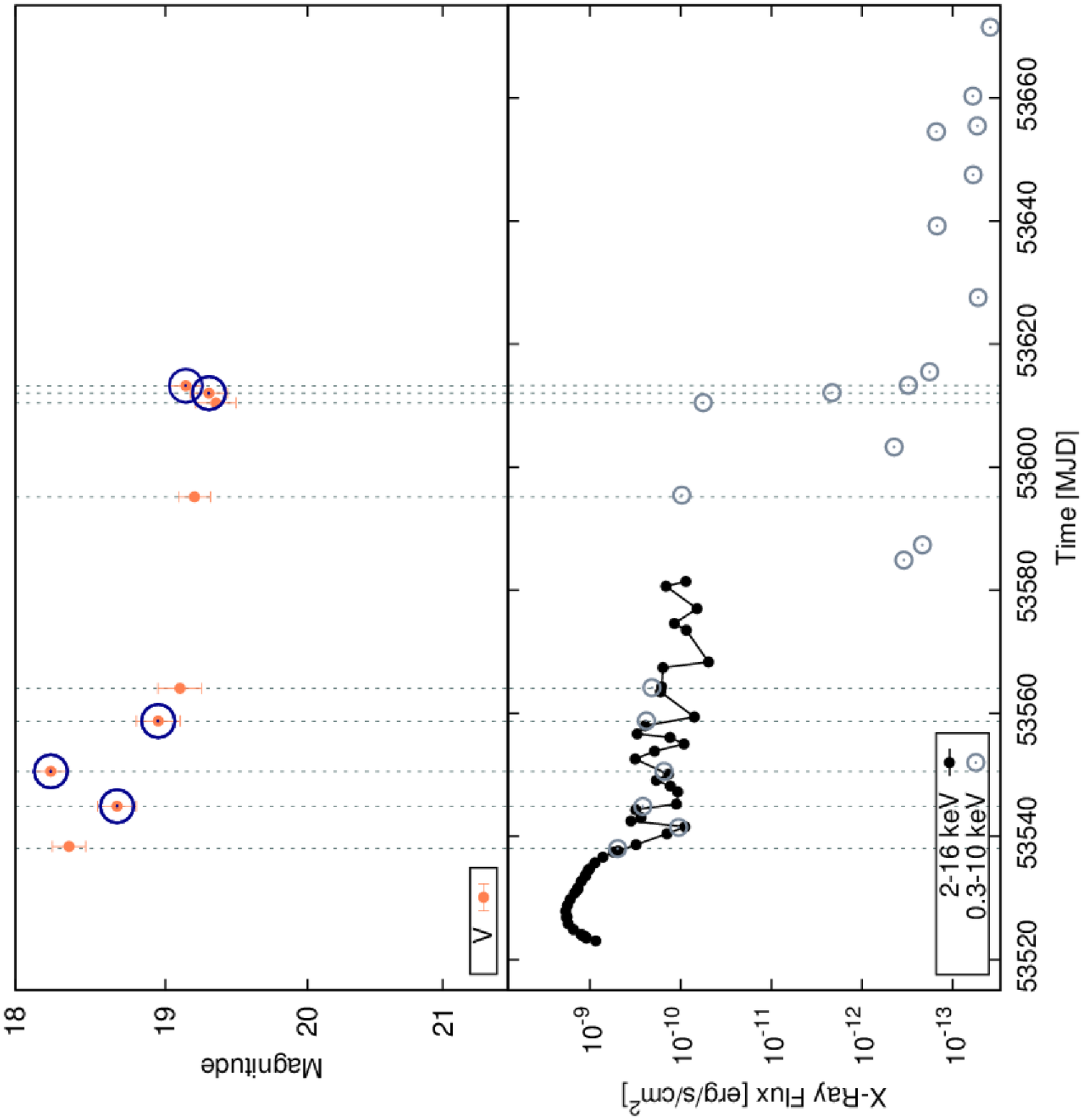}}
  \caption{\textit{Top Panel:} \textit{v} band lightcurve
    (\textit{Swift}/UVOT) of SAX J1808.4--3658 during the 2005
    outburst reflare. The magnitudes are \textit{not} dereddened.
    \textit{Bottom Panel:} 2005 outburst X-ray lightcurve observed
    with \textit{RXTE}/PCA (2--16 keV) and \textit{Swift}/XRT (0.3--10
    keV).}
              \label{fig:2005}%
    \end{figure}

Since the minimum magnitude detected (for the extraction region
centered around \saxj\,) is: $\sim20$ in \textit{v}, $19.4$ in
\textit{b}, ${\sim}18.7$ in \textit{u} and $19.3$ in \textit{w1} we
can confidently exclude that the contaminating source has an important
effect in any but the \textit{v} band.

\section{Results}

\subsection{The Multi-Wavelength Lightcurve}\label{sec:lc}

The multi-wavelength lightcurve of the 2008 outburst is shown in
Figure~\ref{fig:1}.  The reflares are observed at all wavelengths
(with the exception of the first reflare, not monitored by
\textit{Swift}) with a typical cadence of about one observation every 2 days
for \textit{Swift}/XRT and UVOT, every 1-2 days in NIR with
\textit{SMARTS} and multiple observations per day with
\textit{RXTE}/PCA. The main outburst is fully covered with a high
cadence monitoring with \textit{RXTE} and observed by \textit{SMARTS}
for only two days close to the peak of the outburst and by
\textit{Swift} for 10 consecutive days during the slow decay.

The first \textit{Swift}/XRT reflare observation is a non-detection
and corresponds to an upper limit (95\% confidence level) of less than
5--10 times the value detected in quiescence
($L_{q}\simeq5$--$8\times10^{31}\ergs$,~\citealt{cam04b,
  hei07,hei09}).  The timescale of the reflares is about 7--12 days,
which is substantially longer than the fast decay timescale
(${\sim}3$ days; for a detailed discussion of the slow/fast
decay in \saxj\, see~\citealt{har08}). By comparing the
behaviour of the \textit{I} and \textit{H} magnitudes with that of the X-rays it appears
that the NIR luminosity is higher than what would be expected
when extrapolating the behavior seen during the slow
decay. Indeed the two NIR observations occurring right after the
beginning of the fast decay have an \textit{H} and \textit{I} magnitude higher than
what would correspond to the X-ray flux detected on the closest dates
(see Figure~\ref{fig:1}). Therefore it is possible that reflares start
\textit{earlier} in NIR by at least 1.5 days.  The beginning
of the other two reflares is ill constrained since the
sampling is not sufficiently dense in X-rays, but the observations are
still compatible with an earlier response of the NIR.

The last detection of the reflaring phase occurs at 
MJD~54781 ($I$ band) at a level compatible with what
observed during the reflares minima. The source is not detected in the
$H$ band and no observations are performed during that time with the
\textit{Swift}/XRT and UVOT. It is therefore difficult to determine
whether we have detected the beginning of a new reflare, the beginning
of a ``metastable'' state similar to what reported by \citet{cam08} at
the end of the 2005 outburst or something different. The
\textit{I} band magnitude detected at MJD~54781 is ${\sim}2.5$
mag. brighter than the quiescent \textit{I} detected in the 1998
outburst~\citep{cam04} so that we can at least confidently conclude that some
activity is still ongoing at the time of the
observation. We also inspected the publicly available lightcurves
taken with the All Sky Monitor aboard \textit{RXTE} and the Burst
Alert Telescope on \textit{Swift}. We found a few data points close to
the $3\sigma$ detection level, but none of them is statistically
significant given the number of data points (trials) considered (i.e.,
3248 data points). The constraints on the luminosity placed by these two instruments
are not particularly strong since they correspond to
luminosities of the order of a few times $10^{35}\ergs$, which are
values reached only during the brightest portions of the reflares.


The $v$ band magnitudes (not dereddened) of the 2005 outburst are
shown in Figure~\ref{fig:2005} along with the 2--16 keV and 0.3--10
keV X-ray flux. Several observations (highlighted with blue circles in
the figure) show a lack of correlation between the $v$ magnitudes and
the X-ray flux.

\subsection{X-Ray-Optical/UV/NIR correlation}\label{sec:beta}

By looking at Figure~\ref{fig:1} we can see
 how the X-ray luminosity variations co-vary with
the the NIR, optical and UV luminosity. However, the excursion seen
in X-rays spans at least 2 orders of magnitude, whereas the
NIR/optical/UV changes by approximately 3 magnitudes, which translates
into a flux variation of a factor ${\approx}10$--20. Therefore,
we do not expect a steep correlation between X-ray flux and
NIR/optical/UV magnitudes. Furthermore, on several occasions the
X-rays and NIR/optical/UV do not correlate and seem to be varying
independently (see e.g., circled points in Figure~\ref{fig:2005} and
data points around MJD 54760 in Figure~\ref{fig:1} and discussion in 
Section~\ref{sec:mw}).

In this section we estimate the correlation between these
quantities. Since the variations in luminosity can happen on very
short timescales, we decided to consider only the optical, UV and
X-ray emission recorded with the UVOT and the XRT and to exclude the
ground-based optical/NIR data since they are not taken simultaneously
with the X-ray data, but have a typical offsets of ${\sim}0.5$--1 day.
The advantage of using the UVOT and XRT data is that these telescopes
operate strictly simultaneously (although the exposure times are not necessarily
the same).  One exception is represented by one
ground-based $I$ and $H$ band observation from MJD~54758 taken within
1 hour of a \textit{Swift} pointing. In this case we decided to
consider the ground-based data as (quasi)simultaneous.

\begin{figure}
     \centering
     \rotatebox{-90}{\includegraphics[width=0.7\columnwidth]{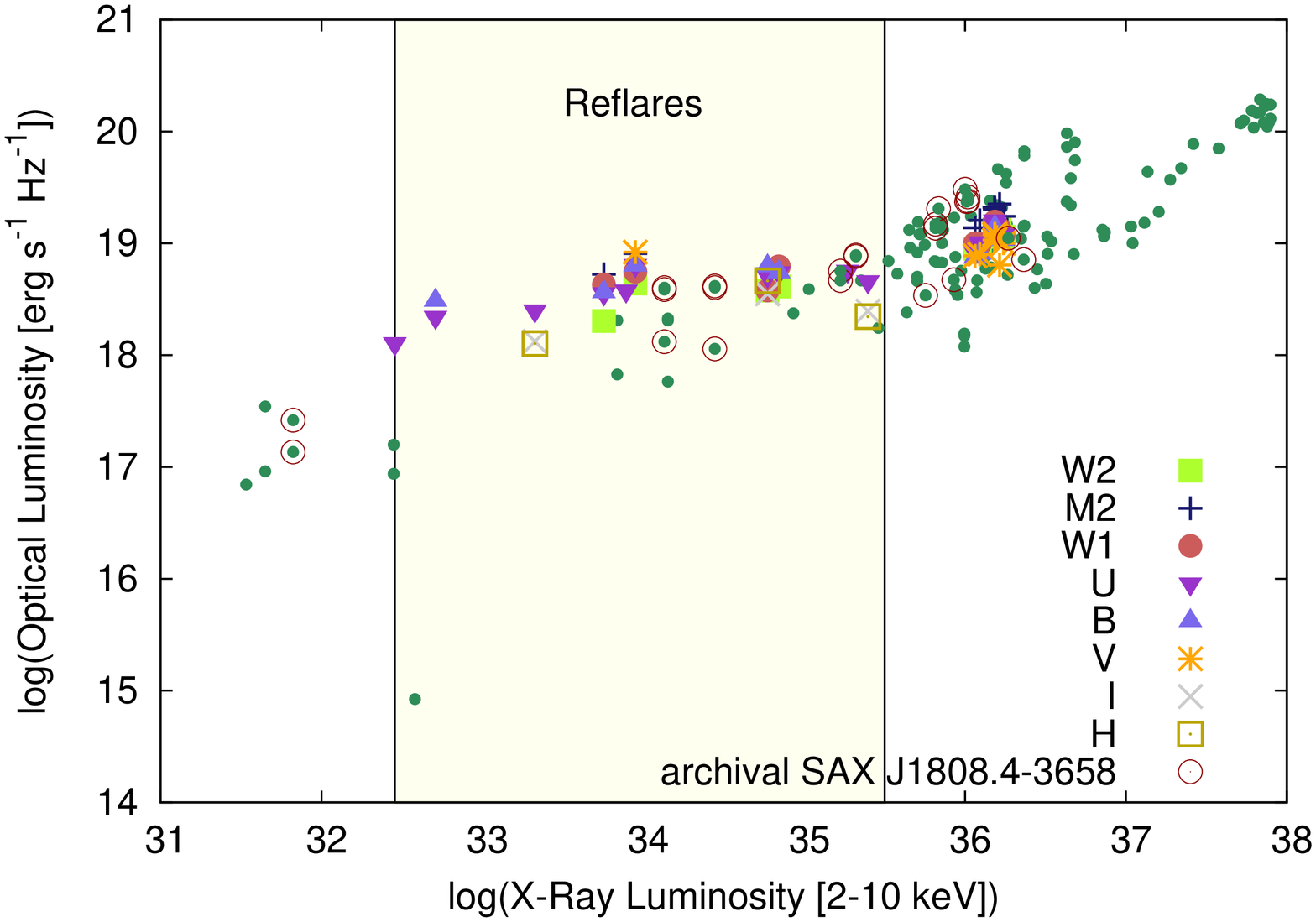}}
     \rotatebox{-90}{\includegraphics[width=0.7\columnwidth]{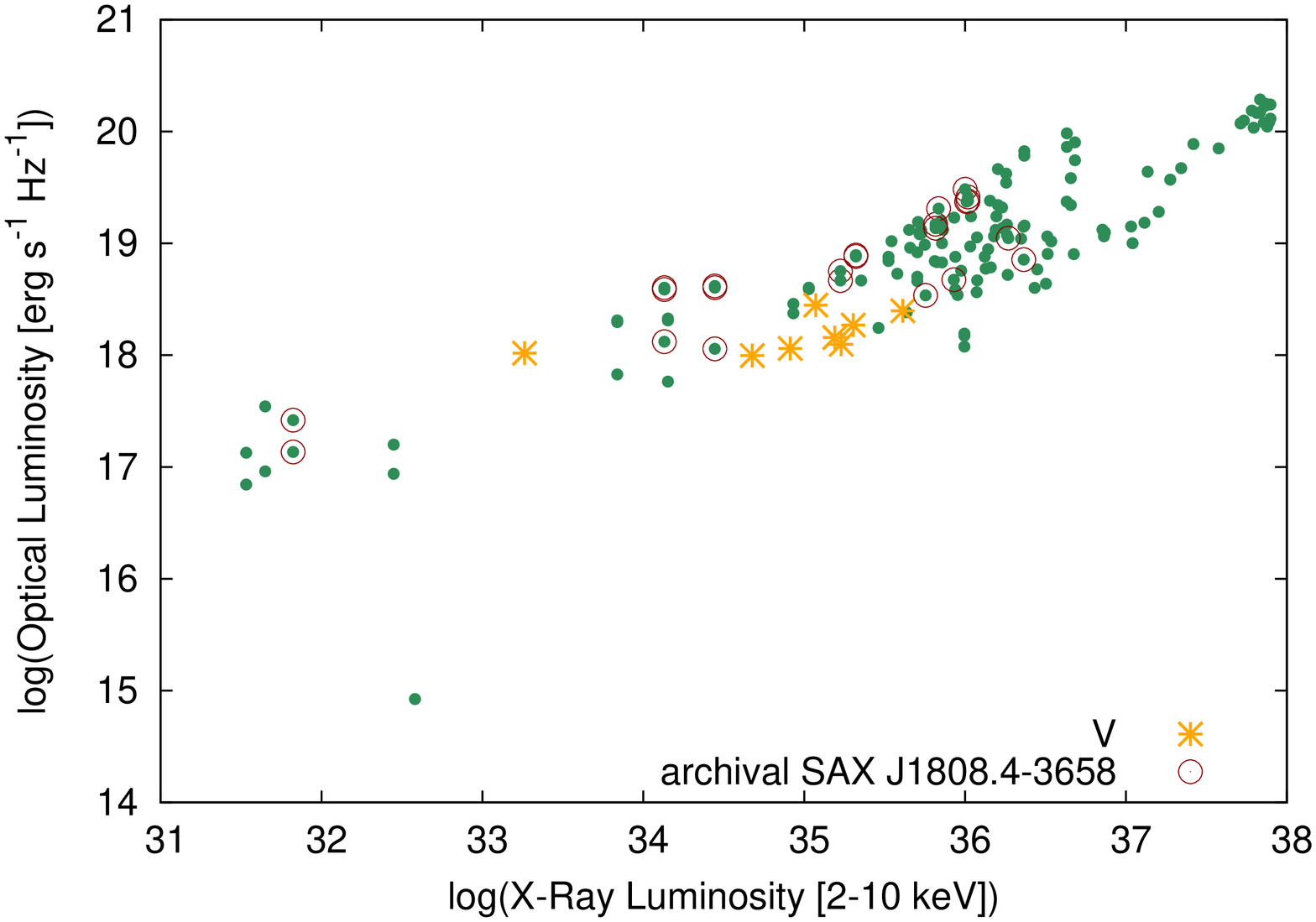}}
   \caption{\textit{Top Panel}: NIR--optical--UV/X-ray correlation diagram. The
     green points show the correlation for several different neutron
     star LMXBs~\citep{rus06,rus07}. A few points (circled in red)
     refer to (archival) SAX J1808.4--3658 data taken during
     quiescence or during its 1998 outburst. The 2008 \textit{Swift}
     observations reported in this paper are shown with different
     symbols/color depending on the specific filter used. \textit{Bottom Panel}: Same as top panel but for the 2005 outburst.}
              \label{fig:2}%
    \end{figure}

In Figure~\ref{fig:2} we plot the de-reddened optical to X-ray
(unabsorbed) luminosity relation for the \textit{Swift} observations
of the 2005-2008 outbursts plus additional data referring to other
neutron star low mass X-ray binaries where these two quantities are
measured (and found to be correlated, see \citealt{rus06,rus07} for
details).  We used the 2--10 keV energy band rather than the 0.3--10
keV in order to match the analysis of~\citet{rus06} done for the other
neutron star LMXBs used in Figure~\ref{fig:2}. Some of the 2008 data
lie slightly above the correlation with a slope that is somewhat
flatter than recorded for other neutron star LMXBs.  The offset
between our data points and the other archival data is not surprising
since the latter points are taken with the $BVRI$ filters, whereas our
data points are mainly composed by $u$ and UV data and only very few
data points are taken with the $I$, $b$ and $v$ filters. Indeed such
effect is not observed in 2005 when the UVOT used only the $v$ filter.
Similar (instrumental) offsets have been observed already in other
sources (see e.g. \citealt{van13}).

\begin{figure}
     \centering
     \rotatebox{-90}{\includegraphics[width=0.7\columnwidth]{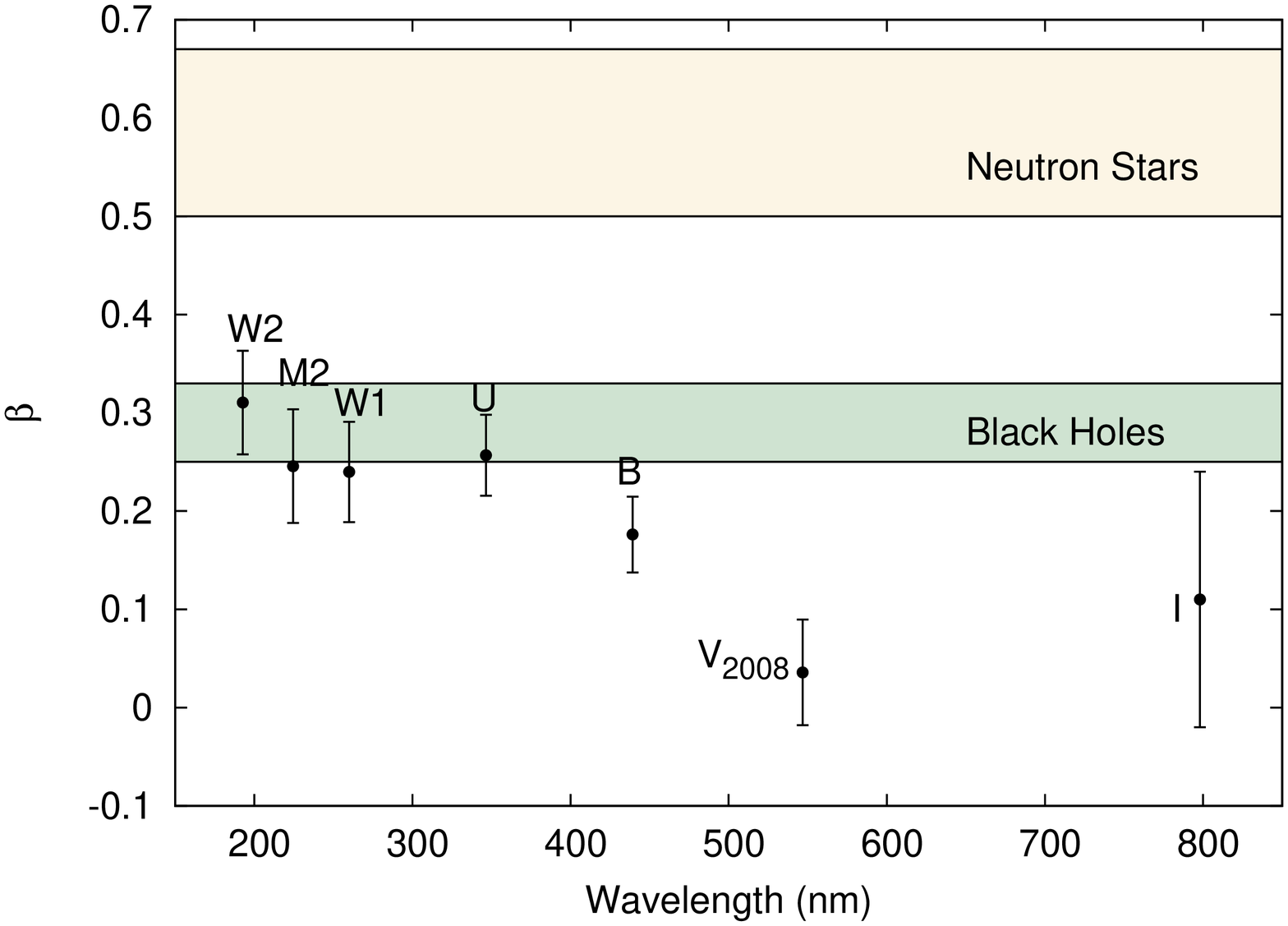}}
   \caption{Correlation coefficient $\beta$ calculated for seven
     different optical/UV bands for the 2008 outburst. The $\beta$
     value for the \textit{v} band of the 2005 outburst is
     unconstrained and is not displayed. The two colored bands
     indicate the theoretical $\beta$ values expected for black hole
     LMXBs (in green) and neutron stars LMXBs (in yellow). The
     difference in the location of the two bands reflects the
     different accretion efficiency ($\propto\dot{M}$ for neutron
     stars and $\propto\dot{M}^{2}$ for black holes). Most data points
     of \saxj\, fall within the black holes band.  The value of
     $\beta$ is also consistent with increasing towards shorter
     wavelengths, a typical behaviour expected in viscous heated
     accretion discs rather than irradiated ones.}
              \label{fig:3}%
    \end{figure}

The flatter slope instead needs an explanation and to quantify this,
we first fitted a simple power law relation
$F_{UV/Opt}=k\,F_{X}^{\beta}$, where $F_{UV/Opt}$ is the flux in one
of the optical/UV bands, $F_{X}$ is the 2--10 keV X-ray flux, $\beta$
is the slope and $k$ is a constant of proportionality.
\citet{rus06,rus07} estimated the coefficient $\beta$ for neutron
stars (assuming a radiatively efficient accretion flow,
$L_X\propto\dot{M}$) and black holes (with a radiatively inefficient
accretion flow, $L_{X}\propto\dot{M}^{2}$).  Since we are using
optical/UV wavebands, the expected range of $\beta$ for a viscously
heated disk is $0.5<\beta<0.67$ for neutron stars and
$0.25<\beta<0.33$ for black holes~\citep{rus06}. In
Table~\ref{tab:beta} and Figure~\ref{fig:3} we report the results of
our fit.
\begin{table}
\begin{center}
\caption[]{Correlation coefficient between the UV/optical and X-ray fluxes}
\label{tab:beta}
\begin{tabular}{ccc}
\hline 
\hline
UVOT/Band& $\lambda_{central}$  (\AA)& $\beta$ \\
\hline  
\textit{w2} &	 1928 & $0.28\pm0.05$\\
\textit{m2} &    2246 & $0.24\pm0.06$\\
\textit{w1} &    2600 & $0.22\pm0.05$\\  
\textit{u}  &    3465 & $0.23\pm0.05$\\	
\textit{b}  &    4392 & $0.16\pm0.04$\\ 
\textit{$v_{2008}$}  &    5468 & $0.03\pm0.05$\\ 
\textit{$v_{2005}$}  &    5468 & $0.05\pm0.21$\\ 
\textit{I}  &    7980 & $0.11\pm0.13$\\
\hline
\end{tabular}
\end{center}
\end{table}

All bands, with the exception of the $v$ and $I$ bands, show a flux that
is clearly correlated with the X-ray flux.  However, the value of the
correlation coefficient $\beta$ falls within the 0.15--0.25 range
(again, excluding the $v$ and $I$ bands). These values are significantly
smaller than those seen in other neutron star LMXBs
(e.g.,~\citealt{arm13,mai08}). The values of $\beta$ also
increase with decreasing wavelength, as found in viscously heated
disks (see e.g. \citealt{rus06, arm13}). For
irradiated disks or for jet-dominated optical emission, the expected
values of $\beta$ are also higher than observed.

\subsection{Rayleigh-Jeans limit}\label{sec:rj}

\begin{figure}
  \centering
  \rotatebox{-90}{\includegraphics[width=1.8\columnwidth]{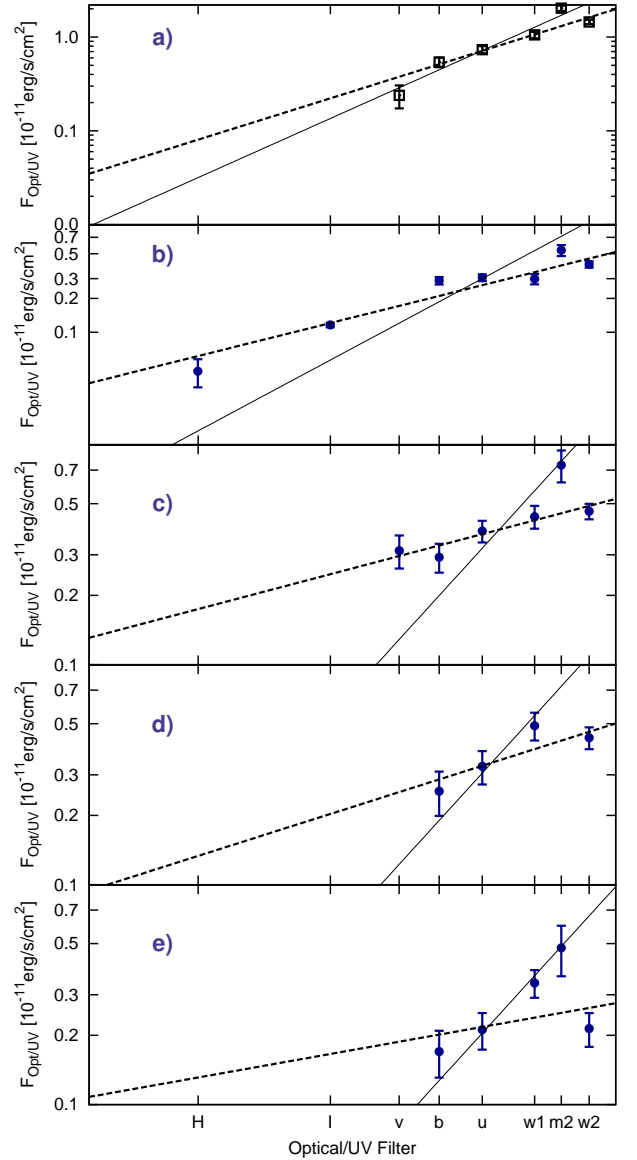}}
  \caption{Optical/UV flux vs. UVOT filter central frequency. In the top panel we show the flux $F_{opt/UV}(\nu)$ corresponding
to one observation taken during the main outburst. The dashed and solid lines represent the best fit to the data 
for $F_{opt/UV}\propto\nu^\alpha$ (with $\alpha$ a free parameter) and $F_{opt/UV}\propto\nu^2$, respectively. The results of the 
fit are reported in Table~\ref{tab:alpha}. }              \label{fig:rj}%
\end{figure}

The shallow slope of the $F_X$ vs. $F_{opt/UV}$ relation could also be
explained if the optical/UV emission falls in the Rayleigh-Jeans
(RJ) limit of the multicolor-blackbody accretion disk spectrum. To test this
hypothesis we selected five observations with the source detected in
at least four UVOT filters. Of these observations, four were taken
during the reflares whereas a ``control'' group was chosen from the
main outburst.  We then fitted $F_{opt/UV} \propto \nu^{\alpha}$
and checked whether $\alpha{\approx}2$ as expected in the RJ limit.
In one observation we included also the ground-based optical/NIR
observations since these data were recorded within 1 hour from the
UVOT and XRT data. We also fitted a correlation with $\alpha$ forced
to be equal to 2 as expected in the RJ limit. In Table~\ref{tab:alpha}
and Figure~\ref{fig:rj} we report the observation selected and the
results of our fit.

\begin{table*}
\begin{center}
\caption[]{Observations Used to test for the Rayleigh-Jeans limit}
\begin{tabular}{ccccccc}
\hline \hline
Panel & Phase & ObsId & MJD & Filters & $\alpha$ & $\chi^2/$dof\\

\hline
a & Main Outburst & 30034031 & 54738.4 & \textit{v,b,u,w1,m2,w2} & $1.4\pm0.3$& 37.7/4\\
b & Reflare & 30034036 & 54758.0 & \textit{H, I, b,u,w1,m2,w2} & $0.93\pm0.11$& 30.5/5\\
c & Reflare & 30034037 & 54760.3 & \textit{v,b,u,w1,m2,w2} & $0.48\pm0.18$ & 7.37/4\\
d & Reflare & 30034038 & 54762.1 & \textit{b,u,w1,w2} & $0.58\pm0.28$ & 2.9/2\\
e & Reflare & 30034039 & 54764.8 & \textit{b,u,w1,m2,w2} & $0.32\pm0.50$ & 10.80/3\\
\hline \hline
\end{tabular}
\label{tab:alpha}
\end{center}
\end{table*}

None of the observations show an $\alpha$ compatible with being 2.
Three out of five observations cannot be fitted with a simple linear
relation (i.e., unacceptable $\chi^2$).  However, by looking at the
panel ``a)'' of Figure~\ref{fig:rj}, which refers to the observation
taken during the main outburst, it appears that we might still be
in the RJ limit for all optical/UV filters if the location of the data
points is somehow affected by unaccounted systematics (see discussion
below).  Panels ``d)'' and ``e)'' are compatible with 
the RJ limit if the peak of the black-body emission falls
around the \textit{m2} and \textit{w2} band. In the panel ``b)'' and
``c)'' there seems to be stronger deviations from the RJ
limit, with $\alpha$ being closer to values of 0--1. These two
observations might be explained if the blackbody peak towards lower
energies.  However, the optical/UV luminosity is slightly lower than
that observed in panels ``d)'' and ``e)'', which is the opposite
effect that one would expect.

An obvious source of systematic uncertainty is the value of $E(B-V)$
(dependent on the value of $N_{H}$) which we have assumed to be
constant throughout the outburst. To verify whether this is indeed
affecting our results we fitted our observations with the model
\texttt{reddening*diskbb} via the software \textit{XSPEC} (v.12.8.1)
by leaving $E(B-V)$ as a free parameter and fixing
$N_H=1.4\times10^{21}\rm\,cm^{-2}$, $T_{in}=0.1$ keV and
$i=50^{\circ}$ (e.g., \citealt{pat09e}). In this way we also take
properly into account the value of reddening as a function of
wavelength over the entire filter bandwidth.  The first two
observations have best-fits that are statistically unacceptable with
null hypothesis probabilities $\ll0.01$. Changing the temperature of
the diskbb model does not change the results in any significant way.
The third and fourth observations are compatible with a value of
$E(B-V)\simeq{0.01}$--$0.16$, which appears to be only marginally
compatible with the galactic $N_H$ \citep{kal05}.  The last two
observations give a range of reddening which is
unconstrained. Therefore we conclude that the data cannot be explained
with a simple multicolor-disk blackbody and that there is no evidence
for the data to be close to the RJ limit and therefore the low values
of $\beta$ need a different explanation.

\begin{figure}
     \centering
     \rotatebox{0}{\includegraphics[width=1.0\columnwidth]{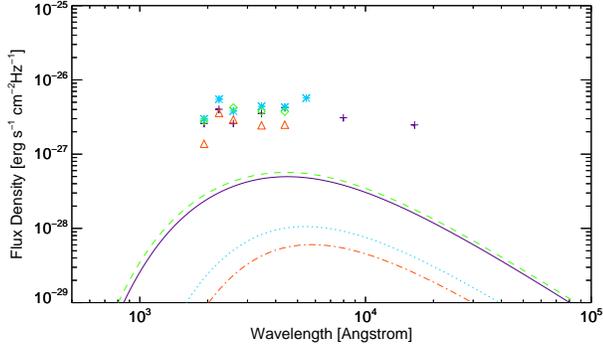}}
   \caption{NIR/Optical/UV emission observed during the reflares (the
     symbols, with colors corresponding to different days, which are
     reported in Table~\ref{tab:alpha}) and and the predicted
     blackbody disk emission for the X-ray luminosity (continuous
     curves, with colors representing different days). Each curve
     corresponds to a specific $\dot{M}$ inferred from the X-ray
     luminosity via the relation $L_{\rm X} \simeq GM\dot{M}/R_*$}
             \label{fig:excess}%
    \end{figure}

Finally, we show that indeed the NIR/optical/UV emission is always
larger (by a factor $\sim10$) than would be predicted assuming that
the X-ray emission tracks the accretion rate so that $L_{\rm X} \simeq
GM\dot{M}/R_*$, and that the outer disk is a standard Shakura-Sunyaev
$\alpha$-disk.  This is seen in Figure \ref{fig:excess}, which plots
the observed NIR/optical/UV flux of the 2008 outburst between MJD
54757 and 54770 and the predicted optical disk emission based on the
X-ray luminosity (although note that this is based on the 2--10 keV
X-ray emission, which could underpredict the bolometric luminosity by
a factor of up to 3~\citealt{int07}). The optical emission is brighter
than would be predicted by X-rays, by a factor of a few 
 (but it would track more closely if the X-ray bolometric flux is higher).

\subsection{X-Ray Spectra and Colors}\label{spec}

\begin{figure}
     \centering
     \rotatebox{-90}{\includegraphics[width=0.72\columnwidth]{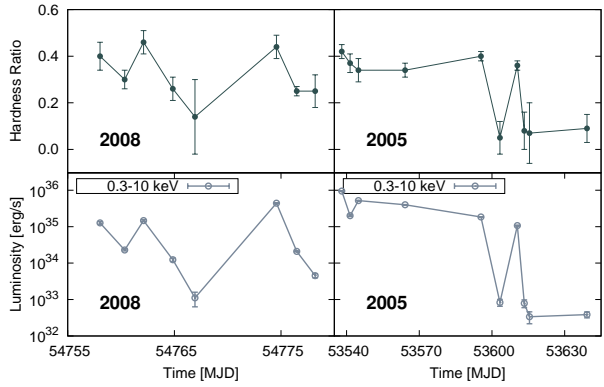}}
   \caption{Hardness ratio (top panels; defined as the ratio between
     the count rate in the 2--10 keV and 0.3--2 keV band) of the 2008
     (left panels) and 2005 (right panels) reflares.  On the bottom
     panels we report the \textit{Swift}/XRT lightcurves for
     comparison (we use only PC mode data and select only observations
     that gave useful constraints on the hardness ratio). The HR and
     X-ray flux are correlated in both outburst: the spectrum becomes
     softer at lower luminosities.}  \label{fig:HR}%
    \end{figure}

We have analyzed the X-ray spectra and performed
a color analysis of SAX J1808.4--3658 during the 2008 and 2005
reflares (with \textit{Swift}) to highlight any possible change in the
accretion flow properties and/or geometry.  In Figure~\ref{fig:HR} we
show the hardness ratio results for the \textit{Swift}/XRT
observations: in both outbursts the HR shows variability which seems
correlated with the variations in luminosity, i.e. the source becomes
softer at lower luminosities.

Applying a standard Spearman's rank test we find an correlation
coefficient of 0.93 and 0.72 (for the 2008 and 2005 outbursts
respectively) but we note that the errorbars associated with the
colors and luminosities are significant and unaccounted for in this
method. Using a composite Monte Carlo analysis of Spearman's rank test
-- as detailed in~\citet{cur14} and implemented in the code {\it
  MCSpearman} (Curran 2014) -- we find a correlation
coefficient of 0.71$\pm$0.21 and 0.66$\pm$0.18 (for the 2008 and
2005, respectively) . This method uses a Monte Carlo re-sampling and
perturbation analysis to return a robust correlation coefficient that
accounts for both the uncertainty associated with the sample and the
errors on individual data points.  We therefore see a significant correlation 
between X-ray luminosity and HR, although we cannot draw any robust
conclusion about the the lowest luminosities. 

In our fits to the \textit{Swift}/XRT spectra from the reflaring portion
of the 2008 outburst (see Table~\ref{tab:spectra}) we used a
spectral model consisting of some combination of a power law
(\texttt{pegpwrlw}), a blackbody (\texttt{bbodyrad}), and a disk
blackbody (\texttt{diskbb}), modified by photoelectric absorption. We
used the \texttt{phabs} absorption model with \texttt{wilm} abundances
\citep{wilms2000} and \texttt{vern} cross sections\citep{verner1996};
the absorption column was in all cases fixed at a value of
$N_\mathrm{H} = 1.4\times10^{21}\textrm{ cm}^{-2}$
\citep[see][]{pat09e}. For our spectral fits and count-rate
conversions of the \textit{Swift} data from the reflaring portion of
the 2005 outburst we used (in all cases) a model consisting of only an
absorbed power law.
The disk blackbody model is required only in the two brightest
observations of 2008.  The choice of a disk blackbody is justified
because such component is also observed during the main outburst (see
e.g., \citealt{pat09e, pap09, kaj11}). The results of our analysis are
reported in Table~\ref{tab:spectra}.

The power law index $\Gamma$ shows little variations and all photon
index values can be well fitted with a simple constant $\Gamma=1.75\pm0.06$
with a $\chi^{2}/dof=6.52/7$. However, given the poor photon counts
and large error bars, the behavior at the lowest luminosities remains
mostly unconstrained.

We detect a blackbody with a small radius, which suggests we are
observing the hot-spot or the neutrons star boundary layer rather than
the thermal emission from the inner accretion disk (see also
~\citealt{pat09e, pap09} for a similar analysis on \textit{XMM-Newton} data
taken during the main outburst).
\begin{table*}
\begin{center}
\caption[]{Spectral Fits to SAX J1808.4--3658 During the 2008 Reflaring State.}
\label{tab:spectra}
\begin{tabular}{llllllll}
\hline 
\hline
DiskBB + BB + PL\\
\hline
Observation & DiskBB Temp. & DiskBB Radius & BB Temp. & BB Radius & $\Gamma$ & $L_{\rm\,0.3-10keV,unabs.}$ & $\chi^2/$dof \\
\hline
& [keV] & [km] & [keV] & [km] & & [$10^{33}\rm\,erg/s$] &\\
\hline
30034041 & $0.155^{+0.012}_{-0.013}$ & $55.0^{+9.4}_{-7.3}$ & $0.485^{+0.018}_{-0.017}$  & $4.75^{+0.47}_{-0.61}$ & $1.42^{+0.22}_{-0.29}$ & $684^{+14}_{-13}$ & 41.2/50\\
\smallskip
30034042 & $0.097^{+0.026}_{-0.023}$ & $130^{+166}_{-58}$ & $0.550^{+0.064}_{-0.052}$  & $2.60^{+0.79}_{-0.58}$ & $1.87^{+0.11}_{-0.16}$ & $490^{+14}_{-14}$ & 55.8/51\\
\hline
BB + PL\\
\hline
\smallskip
30034035\hfill &\hfill&\hfill&  -                      &   -                & $2.00$ (fixed)     & $0.62^{+0.38}_{-0.28}$ & - \\
\smallskip
30034036 & & & 0.50(fixed)             & $2.28^{+0.27}_{-0.31}$   & $1.50^{+0.20}_{-0.26}$ & $105^{+12}_{-12}$ & 7.9/10\\
\smallskip
30034037 & & & $0.40^{+0.09}_{-0.06}$        & $1.27^{+0.58}_{-0.41}$   & $1.88^{+0.23}_{-0.24}$ & $20^{+16}_{-16}$ & 10.2/14\\
\smallskip
30034038 & & & $0.49^{+0.08}_{-0.06}$        & $2.40^{+0.73}_{-0.53}$  & $1.51^{+0.30}_{-0.31}$ & $122^{+13}_{-13}$ & 20.3/18\\
\smallskip
30034039 & & & 0.35(fixed)             & $0.62^{+0.46}_{-0.62}$   & $1.89^{+0.19}_{-0.20}$  & $11.7^{+1.9}_{-1.9}$ & 3.5/5\\
\smallskip
30034040 & & & -                       &                   & $2.00$ (fixed)      & $1.1^{+0.5}_{-0.4}$ & -\\
\smallskip
30034043 & & & $0.35^{+0.03}_{-0.02}$        & $1.95^{+0.34}_{-0.31}$   & $1.73^{+0.14}_{-0.18}$ & $18.6^{+0.9}_{-0.9}$ & 36.4/39\\
\smallskip
30034044 & & & 0.25(fixed)             & $1.16^{+0.50}_{-1.16}$   & $1.75^{+0.29}_{-0.43}$ & $4.3^{+0.7}_{-0.6}$ &  -\\ 
\hline
\hline
\end{tabular}
\end{center}
\end{table*}
The two brightest reflare observations that need a multi-temperature disk blackbody (\texttt{diskbb})
still require a rather cold inner disk temperature (0.1--0.2 keV).  

\saxj\, has remained in the hard state throughout its outburst history
(from 1998 until 2011) with very little spectral variations (see e.g.,
\citealt{van05, har08, bul14}). We caution that the minimum
X-ray luminosity observable with \textit{RXTE}/PCA is of the order of
a few times $10^{34}\ergs$.

\section{Discussion}

In this paper we have presented a multi-wavelength analysis on the accreting millisecond
pulsar \saxj\, that can help to address two specific (and related) issues: 
\begin{itemize}
\item what is the geometry of the accretion flow during reflares?
\item what is the origin of reflares and how are they affected by the presence of a magnetosphere?
\end{itemize}

\subsection{Accretion Flow Geometry}\label{sec:afg}

To understand how the accretion flow behaves during reflares it is
important to consider first whether the inner regions of the
accretion flow are geometrically thin or thick and then how 
the flow interacts with the neutron star magnetosphere.

Since \saxj\, has a measured dipolar magnetic field
(at the poles) of $2\times10^8$ G~\citep{dis08,har08}, it is
possible to estimate the location of its magnetospheric radius. 
We first define the magnetospheric radius $r_{\rm m}$ as the point
where the magnetic field is strong enough to \textit{enforce co-rotation} of
gas in a thin Keplerian disk:
\begin{eqnarray}
r_{\rm m} &=& \left(\frac{\eta\mu^{2}}{4\Omega_*\dot{M}}\right)^{1/5}\\
\nonumber &=& 2.3\times10^6 \eta \left(\frac{B_*}{10^8\rm{~G}}\right)^{2/5}\left(\frac{R_*}{10^6
    \rm{~cm}}\right)^{6/5}\\
\nonumber && \times\left(\frac{P_*}{2\times10^{-3}{\rm
      ~s}}\right)^{1/5}\left(\frac{\dot{M}}{1.6\times10^{-10}
    {~\rm M_{\odot}~yr^{-1}}}\right)^{-1/5}{~\rm cm}.  \label{eq:rm}
\end{eqnarray}
In this equation, $\mu = B_*R_*^3$ is the magnetic moment of the star,
$B_*$ is the neutron star magnetic field at the poles, $R_*$ is the
neutron star radius, $\eta \leq 1$ is a dimensionless parameter
characterizing the strength of the disk/field coupling and reflects
the strength of the toroidal magnetic field induced by the relative
rotation between the disk and dipolar magnetic field, $\Omega_*$
($P_*$) is the star's spin frequency (period) and $\dot{M}$ the mass
accretion rate through the disc.  If we use the known parameters of
\saxj\, and we assume that the observed X-ray luminosity is a good
proxy for the mass accretion rate then at the lowest luminosities
$r_m\simeq 160$ km. This value\footnote{The definition of $r_{\rm m}$
  used in this paper is the same as in \citet{spr93} whereas the
  classical expression of $r_{\rm m}$~\citep{pri72}, obtained by
  equating the gas and magnetic pressures, gives a somewhat larger
  value of about 500 km, which would further strengthen the argument
  presented here.} is significantly larger than the co-rotation radius
$r_{\rm co}=31$ km and it is even larger than the light cylinder
radius $r_{\rm lc}\simeq 120$ km.  The co-rotation radius is defined
as that point in the disk where the Keplerian rotational velocity of
the gas is equal to the rotational velocity of the magnetosphere and
as soon as $r_{\rm m}>r_{\rm co}$ a centrifugal barrier sets in and
matter is assumed to be ejected from the system~\citep{ill75,
  rom04}. Therefore when $r_{\rm m}>r_{\rm co}$ one cannot use
Eq.~\ref{eq:rm} since the $\dot{M}$ inferred from the luminosity might
become a bad indicator of the mass flowing in the accretion flow (for
example because of mass ejection). This means that $r_{\rm m}$ cannot
be as large as 160 km, or at least, if it is, it cannot be inferred
from Eq.~\ref{eq:rm}.
\begin{figure}
  \centering
  \rotatebox{0}{\includegraphics[width=1.0\columnwidth]{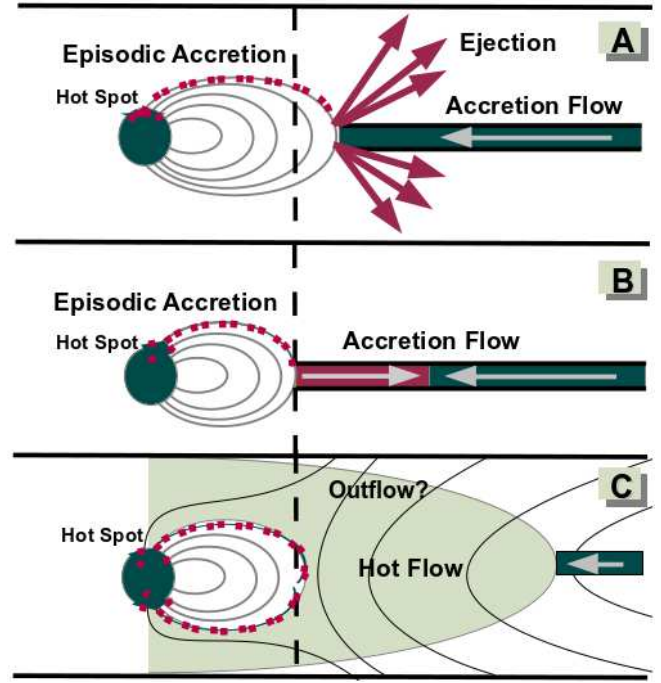}}
  \caption{Different possible accretion states in \saxj\,. The
    vertical dashed line refers to the co-rotation radius $r_{\rm co}$.
    \textit{Panel A.}: strong propeller configuration, when the disk
    is truncated well beyond $r_{\rm co}$ and a very large fraction of
    gas is expelled from the system in an outflow due to the magnetic
    centrifugal barrier. Episodic accretion is still expected to take
    place due to magnetic diffusivity, according to MHD simulations
    (see main text). \textit{Panel B.}: Trapped accretion disk
    scenario. The trapped disk forms when the magnetospheric radius
    reaches $r_{\rm co}$ and the disk remains truncated at that radius
    despite variations in the mass accretion rate. The accretion flow
    is modified in the inner disk regions due to the centrifugal
    barrier, forming a ``dead disk'' configuration (purple segment of
    the disk). The outer disk regions are unaffected by the magnetic
    barrier and the disk is still a standard Shakura \& Sunyaev disk
    (dark green segment). \textit{Panel C.}: Truncated/radiatively
    inefficient accretion flow configuration. The inner disk region is
    set by the a mechanism other than the magnetosphere (e.g.,
    evaporation). The resulting (geometrically thick) hot flow will
    still interact with the magnetosphere possibly generating
    accretion and/or outflows.}
              \label{fig:acc}%
    \end{figure}
If the disk were truncated outside $r_{\rm co}$ it could also be
truncated outside the light cylinder $r_{\rm lc}$. In this case the
magnetosphere should be devoid of matter and the radio pulsar
mechanism should turn on with a strong pulsar wind preventing further
accretion (see e .g.,~\citealt{ste94,bur01}). This is what is
currently thought to occur in the radio-pulsar phase of the three
transitional pulsars recently discovered~\citep{arc09, pap13a, pat14,
  sta14, bas14, roy14} and in the quiescence phase of
\saxj\,\citep{hom01,bur03,cam04}. The fact that in \saxj\, the X-ray
luminosity increases by three orders of magnitude right after reaching
the luminosity minima on a very fast timescale of 1--2 days (see
Figure~\ref{fig:1} and ~\ref{fig:2005}) suggests that the radio pulsar
mechanism does not turn on, although a very rapid switch cannot be excluded 
at the moment. Furthermore, pulsations at high luminosity
(in \saxj\,) and low luminosity (in PSR J1023+0038,~\citealt{arc14},
and XSS J12270-4859,~\citealt{pap14}) argue that accretion onto a
neutron star can take place across a large range of accretion rates 
and therefore accretion onto the neutron star does not need to necessarily 
stop during the lowest luminosity phases of the reflares. 

In the remainder of this paper we will assume that the radio-pulsar
mechanism does not turn on during the reflares (with the
aforementioned caveats) and we will consider three possible scenarios,
sketched in Figure~\ref{fig:acc}, to understand what happens to the
accretion flow during reflares:
\begin{description}
\item[\textbf{A.}] The disk is truncated well beyond the co-rotation radius
  and a strong propeller is ejecting the majority of matter 
flowing in the inner regions of the disk.
\item[\textbf{B.}] The inner disk is close to the co-rotation radius throughout the reflaring phase.
\item[\textbf{C.}] The thin disk is truncated far away from the neutron star
  magnetosphere which is interacting with a geometrically
  thick and radiatively inefficient hot flow.
\end{description}

The first scenario (\textbf{A}) requires that a centrifugal barrier
develops as soon as $r_{\rm m}\gtrsim\,r_{\rm co}$. To enter a
\textit{strong} propeller regime with matter expelled from the system,
$r_{\rm m}$ needs to be larger than $r_{\rm co}$ by at least a factor
${\approx}1.3$ so that the gas can gain enough kinetic energy to reach
the escape velocity~\citep{spr93}.  If a strong
propeller~\citep{ust06, rom05, ill75} is currently operating in the
system then the very-low luminosities we observe are not due to an
extremely low mass flow in the disk, but instead reflect the fact that
only a tiny (poorly-constrained) fraction of the mass actually falls
onto the neutron star surface, generating X-rays. In this case, the
flow itself must likely be radiatively inefficient since otherwise the
much-higher accretion rate in the disc will likely dominate the X-ray
emission. In this specific case the term \textit{radiatively
  inefficient} is not used to indicate the presence of a geometrically
thick radiatively inefficient flow (like an advection dominated
accretion flow) but it is used to describe the fact that the majority
of the accretion energy is not emitted as radiation.

\begin{figure}
  \centering
  \rotatebox{0}{\includegraphics[width=1.1\columnwidth]{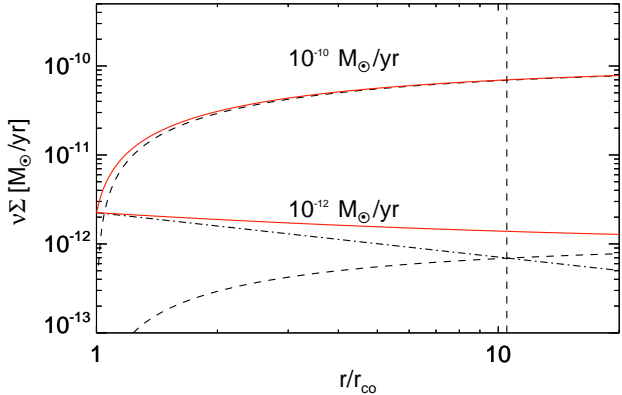}}
  \caption{Surface density profile times kinetic viscosity for an
    accretion disk for two different mass accretion rates. The red
    lines are the total accretion disk density profiles, the dashed lines are
    two standard Shakura \& Sunyaev accretion disk profiles and the dot-dashed
    lines refer to a dead disk density profile. At
    $10^{-12}\rm\,\msun\,yr^{-1}$ the accretion disk is dominated by
    the dead disk solution until $r/r_{\rm co}\approx 10$ and by the
    standard Shakura \& Sunyaev profile beyond that point. At larger
    accretion rates, however, only the regions closest to the
    co-rotation radius are of the dead disk type and most of the
    accretion disk density profile is well described by a standard
    Shakura \& Sunyaev solution. }
              \label{fig:nuSigma}%
    \end{figure}

If $r_{\rm m}$ remains too close to $r_{\rm co}$ so that matter cannot
be ejected from the system, then an accretion disk different than the
standard geometrically thin and optically thick Shakura-Sunyaev type
needs to develop (scenario \textbf{B}). Since the centrifugal barrier
is not sufficient to expel the gas, the angular momentum is
transferred from the magnetosphere to the matter that now remains in
the inner disk regions. The gas piles up and changes the radial
density and temperature profile of that region (see e.g.,
Figure~\ref{fig:nuSigma}). This segment of the disk is then dominated
by a \textit{trapped/dead disk} solution rather than a standard Shakura and
Sunyaev one~\citep{siu77}. In this case the X-ray luminosity is low
because only a small fraction of matter might leak from the inner (dead) disk
towards the neutron star surface \citep{dan10,dan12}.  

    Both these scenarios require that the inner accretion disk radius
    is set by the magnetosphere\footnote{The region over which the
      disk and magnetic field interact will of course have a certain
      radial extent. However, both simulations and theoretical
      arguments predict this transition is sudden and $r_{\rm \rm
        in}=r_{\rm m}$ (see e.g., \citealt{lov95} and an extended
      discussion in \citealt{dan10}).}.  If the innermost region of
    the flow is not a thin disc, but instead forms a
    radiatively-inefficient accretion flow (such as an advection
    dominated accretion flow, ADAF; \citealt{nar95b, nar08, abr95,
      abr00}), then the situation might be somewhat different
    (scenario \textbf{C}). In most radiatively inefficient accretion
    flow (RIAF) models, the accretion flow becomes optically thin but
    geometrically thick and somewhat sub-Keplerian~(see
    e.g.,~\citealt{spr00}).  No detailed study of the interaction
    between a thick disk and a magnetosphere has been undertaken,
    and there is considerably uncertainty about the flow geometry,
    location of the magnetospheric radius and propeller efficiency
    (e.g., \citealt{men01}, \citealt{dal14} and
    \citealt{dan15}). In this case we expect a scenario somewhere between a
    thin-disk/magnetosphere interaction and quasi-spherical accretion
    from a stellar wind (e.g.,~\citealt{sha13}), although there are
    many uncertainties about the properties of such a flow (e.g., the
    thickness of the hot flow, the inner accretion disk location, the
    presence of a jet and outflows; see \citealt{nar08} for a
    review). To assess the validity of this scenario, further
    investigations and theoretical modeling are required.

\subsubsection{Observational Constraints on Accretion Flow Geometry}\label{sec:mw}

Having set the three different configurations for the geometry of the
accretion flow and disk/magnetosphere interaction during the reflares,
we now discuss how the new observations presented in this work and the
past observational evidence collected from the literature can help to
constrain the different scenarios. The three possibilities discussed
above have a number of specific testable features that can be
translated into the following questions:
\begin{itemize}
\item is there a geometrically thin disk extending all the way down 
(or close to) the co-rotation radius? Or is the inner disk radius 
truncated much further away from the neutron star?
\item does the disk/magnetosphere interaction generate a strong propeller
with outflows?
\end{itemize}

\noindent\textbf{Constraints from X-Ray Spectral Analysis.}\\ Our
spectral analysis of the reflares indicates that \saxj\, has an overall
stable spectrum despite the large variations in luminosity.  The
\textit{Swift}/XRT spectra remain hard at all luminosities and all
spectra are consistent with a constant value of $\Gamma\simeq1.7$.  We
do find correlated variations between the \textit{Swift}
hard color and X-ray luminosity that suggests that some (small)
changes in the system are occurring in response to the X-ray
luminosity variations. The \textit{RXTE} colors further show that
\saxj\, is observed in the hard (island) state throughout its eight
outbursts (see e.g., \citealt{int98, van05, har08, bul14} and
Section~\ref{spec}). 
The rather stable X-ray spectrum suggests that the production 
of X-rays is regulated by some stable mechanism despite large
variations in the accretion flow. 

During the brightest portions of the reflares we detect an accretion
disk in addition to a blackbody and a power-law. A very similar
spectrum was seen also during the main 2008 outburst with
\textit{XMM-Newton} \citep{pat09e, pap09, cac09, kaj11}. 

An important caveat is that even if the spectra of the reflares are
similar to the main outburst spectra, this does not necessarily mean
that the accretion flow has the same configuration.  Indeed
the spectra analyzed by \citet{kaj11} and \citet{pat09e},
\citet{pap09} and \citet{cac09} refer to luminosities about one order
of magnitude brighter than the reflares. 

\noindent\textbf{Constraints from X-Ray Aperiodic Variability.}\\ A
truncation radius close to the co-rotation one is also the preferred
scenario to explain the strong 1 Hz modulation observed during the
2000, 2002 and 2005 reflares of \saxj\, (\citealt{pat09c, wij03,
  van00}).  Recently, a similar modulation (1--5 Hz) has been observed
also during the main outburst of the 2008 and 2011 outbursts
(\citealt{bul14}).  In both cases an instability arising from a
trapped disk\citep{dan10,dan12,spr93} has been proposed to be at the
origin of this phenomenon. Such a disk model has also been discussed
to explain a strikingly similar 1 Hz modulation observed in another
accreting millisecond pulsar (NGC 6440 X--2; see \citealt{pat13c}).
It is not possible to exclude that this variability might be related
to propeller driven episodic accretion (e.g., ~\citealt{lii14}) which,
however, needs further investigation.\\

\noindent\textbf{Constraints from Coherent Pulsations.}\\
Even if there are indications for the presence of a thin disk
extending all the way down to the co-rotation radius, we still do not
know whether we are in the presence of a propeller (\textbf{A})
or a trapped disk (\textbf{B}).  Important indications can be given
by the observations of coherent pulsations.

The detection of pulsations during reflares (up to the sensitivity
limit of \textit{RXTE} of about a few $10^{34}\ergs$, see
\citealt{pat09c}) and the presence of a blackbody in the spectrum
(down to luminosities of $10^{33}\ergs$, see Table~\ref{tab:spectra})
does indicate that some gas keeps falling onto the neutron star
surface during most of the reflares. Therefore accretion on the
neutron star surface is not (completely) inhibited during reflares,
most likely at least not down to ${\sim}10^{33}\ergs$.  This is
compatible with the recent findings of accretion powered pulsations in
two quiescent LMXB, where pulsations are seen at
$L_{X}\simeq10^{33}\ergs$~\citep{arc14, pap14}, so that accretion at
these luminosities is not an unprecedented occurrence.  The presence
of pulsations and a blackbody at low luminosities imply that -- if a
strong propeller is currently operating (rather than a ``trapped
disk'') -- sustained accretion onto the neutron star surface is still
ongoing.  Numerical MHD simulations show that during a strong
propeller only a small portion of the mass flow (of the order of few
percent, although the exact value \textit{strongly} depends on the
poorly constrained coupling between the magnetic field and the plasma)
actually reaches the neutron star surface (e.g., via episodic
accretion,~\citealt{lii14, ust06, rom05}) creating a boundary
layer/hot-spot that gives rise to most of the observed X-ray
luminosity. Therefore if a strong propeller is currently operating in
\saxj\, we should expect a very large mass outflow.

Such large outflow of mass might have some further
observational consequences. Indeed if the amount of material reaching
the surface is truly only a very small fraction of the total amount of
matter flung off the disk, then we should still expect a rather strong
spin down of the neutron star.  So far, \saxj\, has not been observed
to spin up or down during its outbursts, with typical upper limits of
$|\dot{\nu}|\lesssim2.5\times10^{-14}\rm\,Hz\,s^{-1}$~\citep{har08,har09,pat12},
a fact that again suggests that this object does indeed stay
close to spin equilibrium ($r_{\rm m}{\simeq}r_{\rm co}$) during most
of its outbursts (see also e.g.,~\citealt{has11}).  
Although these upper limits refer to the entire outburst and not only
to the reflaring portion, we note that at least in one case (the 2000
outburst) \textit{RXTE} collected data exclusively during the reflares
(due to solar constraints). In that case the upper limits on the spin 
frequency derivative were even more stringent: $-1.1\times10^{-14}\rm\,Hz\,s^{-1}\leq\dot{\nu}\leq4.4\times10^{-14}\rm\,Hz\,s^{-1}$ (95\% confidence interval, 
see Table 4 in \citealt{har08}).

Episodic accretion might possibly have a role to explain why we do not
see a strong spin down despite the strong propeller since the angular
momentum loss due to the centrifugal barrier might be (partially)
compensated by the material torque accreting onto the neutron star
surface. However, to determine the order of magnitude expected we can
use the relation:
\begin{eqnarray}
  \dot{J}_{\rm prop} &=& -n \dot{M}_{\rm ej} (GMr_{\rm in})^{1/2} \nonumber\\
               &=& -n (r_{\rm in}/r_{\rm co})^{1/2}
                   \dot{M}_{\rm ej} (GM r_{\rm co})^{1/2} \,,
\end{eqnarray}
where $r_{\rm in}$ is the inner disk radius and the dimensionless
torque $n$ is zero for $r_{\rm in}=r_{\rm co}$ and of order unity for
$r_{\rm in}\gtrsim 1.1\; r_{\rm co}$ \citep{eks05}.  For $r_{\rm
  in}\approx 1$--$2\,r_{\rm co}$ and $\dot{M}_{\rm ej}\approx
10^{-10}\rm\,\msun\,yr^{-1}$ we obtain a spin down of the order of a
few times $10^{-14}\rm\,Hz\,s^{-1}$. Such values are of the same order
of magnitude as the upper limits placed in most of the outbursts (see
e.g., \citealt{har09} and Table 4 in \citealt{har08}).  The choice of
$\dot{M}_{\rm ej}\approx10^{-10}\rm\,\msun\,yr^{-1}$ derives from the
fact that this is the approximate value required to produce X-ray
luminosities of a few times $10^{35}\ergs$, which is the typical value
observed at the peak of the reflares.  Since the material falling on
the neutron star has to be at least $10^{-10}\rm\,\msun\,yr^{-1}$,
then the expected spin-down above is the minimum that we can expect in
this scenario. The constraints on the observations (at least for the
2000 reflares) suggest that any spin down present in \saxj\, must be
smaller than this minimal expected $\dot{\nu}$ value. However, given
the uncertainties in the model and the close value between the
expected $\dot{\nu}$ and the observed upper limits, this result should be
taken more as a possible indication against the strong propeller
rather than conclusive evidence.

The lack of a detected spin-down can also be important to place
constraints on the trapped disk scenario (\textbf{B}). The sign of
the torque depends on whether the inner edge of the disk is inside or
outside the co-rotation region, and the amplitude is roughly determined
by the ``critical accretion rate'' -- $\dot{M}_{\rm c}$, the accretion
rate at which the inner edge of the disk is equal to the co-rotation
radius. In \citealt{dan12} this was estimated as:
\begin{equation}
\dot{M}_{\rm c} = \frac{\eta \mu^2}{4\Omega_*r_{\rm c}^5},
\end{equation}
and corresponds to $\dot{M}_{\rm c} \sim 5\times10^{-11}$M$_{\odot}$
yr$^{-1}$ for \saxj\, (assuming a magnetic field of 10$^{8}$G). The
amplitude of the spin-up/spin-down torque is then roughly:
\begin{equation}
\dot{J}_{\rm c} = \dot{M}_{\rm c}\left(GM_*r_{\rm c}\right)^{1/2}
\end{equation}
which (assuming a moment of inertia of $10^{45}$g cm$^{2}$)
corresponds to a spin change of $\Delta \nu \sim 10^{-14}\rm\,Hz$. For
a mean outburst accretion rate of $\dot{M}\sim2\times10^{-10}$
M$_\odot$ yr$^{-1}$ this corresponds to a spin up rate of $\sim
(0.3$--$1)\times10^{-13}\rm\,Hz\,s^{-1}$.  For the long-term average
accretion rate of $\dot{M}\sim 5\times10^{-12}$ M$_\odot$ yr$^{-1}$
(i.e., averaged over an entire outburst/quiescence cycle), the spin-up
from accretion is nearly balanced by spin-down from the
disk-magnetosphere coupling outside $r_{\rm c}$, and more detailed
calculations predict a net spin down rate of $(1$--$3)\times10^{-14}\rm\,Hz\,s^{-1}$ (see also Figure 2 of \citealt{dan12}).
Such values are not much different than those obtained for the
propeller scenario (\textbf{A}) and indeed similar conclusions can be
drawn: the expected spin up/down from a trapped disk is too close to
the existing measured upper limits to place robust constraints on the
operating mechanism.

Finally, it is worth noticing that the lack of a spin-down during the
outbursts might still be compatible with the scenario \textbf{C},
where a thick hot flow is present up to a very large distance from the
neutron star. In this case it might be the sub-Keplerian nature of the
hot flow itself that results in a very small amount of angular
momentum transferred towards the neutron star and thus in a lack of
detected accretion torques. In Table~6 we summarize
which of the observed properties can be explained by each of the models
considered here. \\

\begin{table*}
  \begin{threeparttable}
    \caption{Models for the Geometric Configuration of the Accretion Flow}
    \centering
    \scriptsize
    \begin{tabular}{lcccccc}
      \hline
      \hline
      Model & X-Ray Spectra & Spin-Down &1 Hz QPO & Weak Irradiation & X-Ray Delay & NIR/Optical/UV Excess\\
      \startdata
      & & & & & & \\
      Strong Propeller (\textbf{A.}) &  P & P & P & ? & Y & Y\\
      \smallskip
      Trapped/Dead Disk (\textbf{B.}) & Y & P & P & ? & Y & Y\\
      \smallskip
      Thick Hot Flow    (\textbf{C.}) & N & P & N & ? & Y & Y\\
      \hline
      \hline
    \end{tabular}
      The table compares observed properties of the to the various
      models discussed in \S 4.  {\bf Y}/{\bf N} (yes/no) indicates
      that the model can/cannot explain the property in \saxj\,.  The
      symbol {\bf P} (possible) indicates that the model might be able
      to explain the property if certain conditions are met.  The
      symbol {\bf ?}  indicates that the model makes no specific
      prediction for that property, and that further studies are
      required.
  \end{threeparttable}
\end{table*}\label{tab:summary}

\subsection{The Origin of Reflares}
Reflares are an important phenomenon to study because they are assumed
to be generated by the ionization instability that drives dwarf novae
and X-ray binary outbursts.  Therefore understanding why reflares are
sometimes observed can give important information about how the
ionization instability operates. Even more importantly, their
observation can help testing the assumption that they are indeed
generated by the ionization instability.

Reflares are currently considered a problem for the disk instability
model since they require that a large amount of matter is retained in
the disk a the end of the main outburst~\citep{ara09, dub01}.  The
currently best explanation (which qualitatively agrees with the
observed timescales and luminosity fluctuations) is that reflares are
essentially `mini-outbursts'. A small change in disk density at the
end of an outburst increases the outer disk temperature enough to
partially ionize hydrogen, which then leads to a rapid rise in the
accretion rate into the inner disk and a rebrightening. The exact
trigger of the reflare is uncertain -- they appear spontaneously in
the simulations of \citet{dub01}, although they do not resemble the
observed reflares and they are seen by \citet{ham00} where reflares
are caused by an increased irradiation of the donor star that causes a
superoutburst (so called because their duration is much larger than
that of normal outbursts) followed by reflares. The donor star in
\saxj\, is observed to be strongly irradiated during
quiescence~\citep{hom01,bur03,cam04} and indeed there are suggestions
that it is losing a large amount of mass (\citealt{dis08}; see however
\citealt{pat12} and \citealt{har08} for criticisms of this strong mass
loss scenario in \saxj\,).  Another suggestion for the reflare trigger
comes from the mass reservoir model of \citet{osa01}, where reflares
are triggered also after superoutbursts as long as the effective
viscosity of the disk (parametrized by the $\alpha$ parameter) remains
large through the entire sequence of reflares.

So far little work has been done on reflares occurring in X-ray
binaries and most theoretical work has focused on explain the reflaring
phenomenon observed in the dwarf novae population. 
However, regardless of the trigger (which might be of a different 
nature than outlined above), the reflares should proceed in the
same way: at the end of an outburst the temperature at some location
in the outer disk falls below the hydrogen ionization temperature,
causing hydrogen to recombine and the accretion rate to fall
dramatically as a `cooling wave' travels inward. For (possibly) one of
the reasons outlined by the irradiation or mass reservoir model, the
density in the disk subsequently rises again enough to reach the
hydrogen ionization temperature and again send a new `heating' wave
through the disk, which causes the accretion rate to rise. Since there
is not a large amount of gas left in the accretion disk after the main
outburst, the accretion rate again decreases, resetting the entire
process.

In WZ-Sae type systems, the reflares are observed after
the occurrence of a superoutburst, so called because its duration is
much larger than that of normal outbursts. In these systems, it was
proposed that a resonance occurs when the disk radius exceeds the
threshold of 0.46$a$ where $a$ is the orbital semi-major axis.  When
this happens superhumps are expected to develop~\citep{osa89, osa96}.
Black hole candidates in X-ray binaries
displaying superhumps do not show outbursts of varied amplitudes~\citep{mac14}
and \saxj\, does follow the same behaviour. It is possible that both \saxj\,
and those black hole candidates have shown only
superoutbursts.~\citet{mac13} suggested indeed that we do observe only
super-outbursts from some LMXBs and superhumps should be detected if
the reflares in \saxj\, are a variant of those seen in dwarf novae.

\citet{ele09} reported weak evidence for the presence of a superhump
at optical wavelengths in SAX J1808.4--3658, which, if confirmed,
might be an exciting observational diagnostic and might indicate the
presence of an eccentric disk. However, the candidate superhump has
been observed so far only once during the main 2008 outburst and it
attends further confirmation.

Regardless of the presence of superoutbursts, this basic hot/cold wave
scenario can plausibly account for the reflares in SAX
J1808.4-3658. If the instability originates in the outer part of the
disk, the reflare rise time can be associated with the time it takes
for the heating front to propagate from the outer edge of the disk to
the central regions, which will happen on a thermal timescale:
\begin{equation}
t_{\rm   r} \sim r_{\rm d}/\alpha c_{\rm s}
\end{equation}
where $r_{\rm d}$ is the location of the outer edge of the disk,
$\alpha = 0.1$ is the Shakura-Sunyaev viscosity parameter, and $c_{\rm
  s}$ is the sound speed at the inner edge of the heating front, where
$T \sim 6500$~K. For a 2-day rise time, $r_{\rm d} \sim 10^{10}$~cm,
which is comparable to the circularization radius of the disk $r_{\rm
  cr}\sim 0.46 a\sim 3\times 10^{10}$~cm (assumed to be the outer disk
edge; $a$ is the projected semi-major axis of the binary, for the
orbital period 7249s and assumed NS mass of 1.4$M_\odot$ and donor
companion of $0.07\msun$).

\subsubsection{Constraints from Multiwavelength Photometry}
During the reflaring phase of the 2008 (and to some extent also 2005)
outburst, we can place three constraints on the
nature of the accretion disk in \saxj\,. 
The first constraint comes form the observation of \textit{weak}
irradiation as witnessed by the small value of $\beta$ (see
Section~\ref{sec:beta}). This usually suggests that the accretion disk
is heated mostly by viscous dissipation rather than irradiation (see
e.g., \citealt{rus06}). The reason why irradiation is not the main
contributor to the optical/UV flux emission remains an open question.

The second constraint comes from the observation of some
anti-correlations between X-ray and optical/NIR emission (e.g.,
Figure~\ref{fig:1} on MJD 54746--54747 and MJD 54758 and several data
point in Figure~\ref{fig:2005}). As discussed in Section~\ref{sec:lc}
there is a tentative X-ray delay of about 1.5 days with respect to the
NIR/optical emission\footnote{We also note that during \textit{some
    observations} of the 2000 outburst (see~\citealt{wac00,wij06}) the
  optical luminosity also did not correlate with the X-ray
  luminosity.}. The X-ray delay is a phenomenon that has been
already observed in other LMXBs although so far only in systems
containing black holes. In some systems (two of which are persistent
sources) the optical/X-ray cross-correlations shows evidence for X-ray
delays on the order of ${\sim}5$--$20$ days. \citet{bro01} found an
X-ray lag of 5-–10 days from long-term optical and X-ray light-curves
of the black hole binary LMC X--3 (a persistent source; see also
\citealt{ste14} who measured this delay as 2 weeks). \citet{hom05}
found an X-ray lag of 15-–20 days compared to near-IR variations in
the black hole LMXB GX 339-–4 (a transient source). An X-ray lag of
2--14 days was observed in 4U 1957+11 (persistent) behind
optical~\citep{rus10}, whereas a 3.5 d X-ray time lag compared to UV
was found in Swift J1910.2-0546 (a transient source; \citealt{nak14}).
Such delays have been interpreted with a variety of models that
propose that the NIR/optical/UV emission originates from a jet, from
the outer disk or even from the irradiated stellar surface.

In the case of \saxj\, we do not have sufficient observational
evidence to constrain the presence of a jet during the reflares.
Since the jet model (or an ADAF with an outflow) cannot be constrained
we will not consider it any further, although we stress that there is
still no proof that can firmly exclude such scenario. 

If we consider a disk origin of the emission then we can assume that
the NIR/optical/UV emission is coming from the outermost disk regions.
In this case by looking at Figure~\ref{fig:excess} it is clear that
this emission is over-luminous with respect to the X-rays
luminosity. It is possible that, when the heating front moves inward
and reaches the inner disk regions, a larger flow of X-rays is
produced which would correspond to the peak of the reflares. However,
the peaks are still too dim for the optical luminosity observed and
therefore this scenario is unable to explain the observed behaviour.

On the other hand, if an outflow is present and a significant portion
of mass is lost from the system then a strongly over-luminous
NIR/optical/UV emission is expected.  Indeed if the material
generating radiation is lost in the disk before reaching the innermost
regions then the X-rays will not be produced. This can explain the
possible X-ray delay, as well as our third important constraint: any
multicolor-disk model of a viscously heated accretion disk seems to
modestly under-predict the optical luminosity observed for the
measured X-ray luminosity (see Section~\ref{sec:rj}). Since
the optical emission is only over-predicted by a factor of a few with
respect to the X-ray flux (assuming a constant accretion rate through
the disk) our results also exclude a very strong outflow in the
disk. Such an outflow would expel the majority of gas before it
radiated energy in X-rays, which would make the discrepancy between
optical and X-rays much larger.
However, it is still possible that some (not too strong) outflow of material
is present in the system, which could be driven either by a
magnetospheric RIAF (like the strong propeller; \citealt{ust06}) or by
an accretion driven RIAF (like a wind; \citealt{bla99}), is present in
the system.

Finally, almost identical conclusions can be reached for the scenario
\textbf{B} (trapped disk).  In Figure~\ref{fig:nuSigma} we compare the
surface density profiles (times the kinetic viscosity) of a standard
accretion disk and a trapped/dead disk. The extent of the trapped disk
region could be modest which means that the largest majority of the
accretion disk would still look and behave as a standard Shakura \&
Sunyaev accretion disk.  This means that at large accretion rates the
total disk profile would be observationally indistinguishable (in
luminosity and spectral properties) from a Shakura \& Sunyaev disk,
despite the presence of a small segment where the dead disk is still
present.  Therefore identical considerations discussed above apply to
this scenario too. We caution that the extent of the region where the
dead disk dominates strongly depends on the efficiency of the angular
momentum extraction at the inner edge of the disk, a quantity that is
currently poorly known.

Interpreting the reflares in terms of a specific physical
model is hampered by the uncertain origin of the UV/optical/NIR
emission. Assuming that it is dominated by a single physical
component, the simplest explanation would be that it originates in the outer
parts of the accretion disk, either directly from viscous dissipation
or as reprocessed emission from irradiation from the inner X-ray
component. However, as discussed in~\ref{sec:rj} and~\ref{sec:mw}, 
we consider it unlikely to be reprocessed emission and its origin 
remains still to be identified.

\section{Conclusions}

In this paper we have analyzed the 2008 and 2005 reflaring phase of
the accreting millisecond pulsar \saxj\,. We find that the accretion
flow geometry can be substantially constrained by the occurrence of
reflares. We find that a disk extending down to the co-rotation radius
and truncated by the neutron star magnetosphere provides the most
likely explanation for the accretion flow.  We have found that the
NIR/optical/UV emission during the reflares is over-luminous and
cannot be explained by an irradiated accretion disk model if the
mass accretion rate used is inferred from the observed X-ray
luminosity.  We suggest that most of the NIR/optical/UV emission comes
from the outer disk and it is produced by viscous dissipation. The
mass accretion rate in the outer disk regions needs to be
substantially higher than the one that reaches the inner disk regions.
We propose that either a propeller with a large amount of matter being
expelled from the system or a trapped (dead) disk truncated at the
co-rotation radius are both plausible explanations for all of these
features. A disk truncated (by e.g., evaporation) far away from
the neutron stars (at hundreds or even thousands gravitational radii) is 
instead not supported by the current observational evidence. 
Finally, we find that a hot/cold wave propagation model for 
the reflarings is compatible with the observed timescales although the 
mechanism that triggers such phenomenon is still to be identified.

\begin{acknowledgements}
We would like to thank Jari Kajava for interesting discussions and
suggestions.  AP acknowledge support from a Netherlands Organization
for Scientific Research (NWO) Vidi Fellowship. This work was partially
supported by Australian Research Council grant DP120102393. This
research has made use of data obtained by the CTIO 1.3m telescope
operated by the \textit{SMARTS} consortium. This work made use of data
supplied by the UK Swift Science Data Centre at the University of
Leicester.
\end{acknowledgements}


\end{document}